\documentclass[a4paper,12pt]{article}

\usepackage{amsmath,amssymb}
\usepackage{cite}

\begin{document}

\author{Nikolay A. Kudryashov\footnote{E-mail: kudryashov@mephi.ru}}

\title{Seven common errors in finding exact solutions of nonlinear differential equations}

\date{Department of Applied Mathematics, \\ National  Research Nuclear University MEPhI,  \\ 31 Kashirskoe Shosse, 115409, Moscow, \\ Russian Federation}

\maketitle


\begin{abstract}

We analyze the common errors of the recent papers in which the solitary wave solutions of nonlinear differential equations are presented. Seven common errors are formulated and classified. These errors are illustrated by using multiple examples of the common errors from the recent publications. We show that many popular methods in finding of the exact solutions are equivalent each other. We demonstrate that some authors look for the solitary wave solutions of nonlinear ordinary differential equations and do not take into account the well - known general solutions of these equations. We illustrate several cases when authors present some functions for describing solutions but do not use arbitrary constants. As this fact takes place the redundant solutions of differential equations are found. A few examples of incorrect solutions by some authors are presented. Several other errors in finding the exact solutions of nonlinear differential equations are also discussed.

\end{abstract}

\newcommand {\psps} {\frac{\psi'}{\psi}}
\newcommand {\Dpsps} {\left( \frac{\psi'}{\psi} \right)'}
\newcommand {\DDpsps} {\frac{\psi''}{\psi}}
\newcommand {\DDDpsps} {\frac{\psi'''}{\psi}}
\newcommand {\pspsPP} {\left( \frac{\psi'}{\psi} \right)^2}
\newcommand {\pspsPPP} {\left( \frac{\psi'}{\psi} \right)^3}
\newcommand {\pspsPPPP} {\left( \frac{\psi'}{\psi} \right)^4}
\newcommand {\pspsDD} {\left( \frac{\psi'}{\psi} \right)''}
\newcommand {\DpspsPP} {\left( \left( \frac{\psi'}{\psi} \right)' \right)^2}
\newcommand {\kz} {\frac{z}{2} \sqrt{4c-1}}
\newcommand {\kw} {kx + \omega t}

\section{Introduction}

During the last forty years we have been observing many publications
presenting the exact solutions of nonlinear evolution equations. The
emergence of these publications results from the fact that there are
a lot of applications of nonlinear differential equations describing
different processes in many scientific areas.

The start of this science area  was given in the famous work by
Zabusky and Kruskal \cite{Zabusky}. These authors showed that there
are solitory waves with the property of the elastic particles in
simple mathematical model. The result of the study of the Korteweg
--- de Vries equation was the discovery of the inverse scattering
transform to solve the Cauchy problem for the integrable nonlinear
differential equations \cite{Gardner}. Now the inverse scattering
transform is used to find the solution of the Cauchy problem for
many nonlinear evolution equations \cite{Lax, Ablowitz, Ablowitz01}.
Later Hirota suggested the direct method \cite{Hirota71} now called
the Hirota method, that allows us to look for the solitary wave
solutions and rational solutions for the exact solvable nonlinear
differential equations \cite{Kudryashov, Polyanin01}.

As to nonintegrable nonlinear evolution equations we cannot point
out the best method to look for the exact solutions of nonlinear
differential equations. However we prefer to use the truncated
expansion method by Weiss, Tabor and Carnevalle \cite{Weiss} and the
simplest equation method in finding the exact solutions. In section
1 we demonstrate that many popular methods to look for the exact
solutions of nonlinear differential equations are based on the
truncated expansion method. Many methods are obtained as consequence
of the truncated expansion method and we are going to illustrate
this fact in this paper.

Nowadays there are a lot of computer software programs like
MATHEMATICA and MAPLE. Using these codes it is possible to have
complicated analytical calculations to search for the different
forms of the solutions for the nonlinear evolution equations and
many authors use the computer codes to look for the exact solutions.
However using the computer programs many investigators do not take
into account some important properties of the differential
equations. Therefore some authors obtain "new" cumbersome exact
solutions of the nonlinear differential equations with some errors
and mistakes.

The aim of this paper is to classify and to demonstrate some common
errors that we have observed studying many publications in the last
years. Using some examples of the nonlinear differential equations
from the recent publications we illustrate these common errors.

The outline of this paper is as follows. In section 2 we present
some popular methods to search for the exact solutions of
nonintegrable differential equations and we show that in essence all
these approaches are equivalent.  In section 3 we analyze some
reductions of partial differential equations to nonlinear ordinary
differential equations. We demonstrate that many reductions have the
well - known general solutions and there is no need to study the
solitary wave solutions for these mathematical models. In section 4
we give some examples when the authors remove the constants of
integration and lose some exact solutions.  In section 5 we
demonstrate that many publications contain the redundant expressions
for the exact solutions and these expressions can be simplified by
taking arbitrary constants into account. In section 6 we discuss
that some solitary wave solutions can be simplified by the authors.
In section 7 we point out that we have to check some exact solutions
of nonlinear differential equations because in number of cases we
have "solutions" which do not satisfy the equations studied. In
section 8 we touch the solutions with redundant arbitrary constants.
For many cases the redundant arbitrary constants do not lead to
erroneous solutions but we believe that investigators have to take
these facts into account.

\section{First error: some authors use equivalent methods to find exact solutions}

In this section we start to discuss common errors to search for the
exact solutions of nonlinear differential equations. We observed
these errors by studying many papers in the last years.

Many authors try to introduce "new methods" to look for "new
solutions" of nonlinear differential equations and many
investigators hope that using different approaches they can find new
solutions. However analyzing numerous applications of many methods
we discover that many of them are equivalent each other and in many
cases it is impossible to obtain something new.

So the first error can be formulated as follows.

\emph{\textbf{First error}.  Some authors use the equivalent methods
to find the exact solutions of nonlinear differential equations, but
believe that they can find new exact solutions}.

Using the solitary wave solutions of the KdV - Burgers equation let
us show that many methods to look for the exact solutions of
nonlinear differential equations are equivalent.

\emph{\textbf{Example 1a.}} \emph{Application of the truncated
expansion method.}

The truncated expansion method was introduced in \cite{Weiss,
Weiss01} and developed in many papers to obtain the Lax pairs, the
Backlund transformations and the rational solutions for the
integrable equations \cite{Conte01}. This approach was also used in
the papers \cite{Kudryashov88, Conte89, Kudryashov90, Kudryashov90a,
Kudryashov91, Kudryashov92, Kudryashov93, Kudryashov94,
Kudryashov96, Peng01} to search for the exact solutions of
nonintegrable differential equations.

Consider the application of the truncated expansion method in
finding the exact solutions of the KdV - Burgers equation

\begin{equation}
\label{SMM}u_t+u\,u_x+\beta\,u_{xxx}=\mu\,u_{xx}.
\end{equation}

Substituting $u(x,t)$ in the form \cite{Weiss}
\begin{equation}
\label{SMMa}u(x,t)=\frac{u_0(x,t)}{F^p}+\frac{u_1(x,t)}{F^{p-1}}+...+u_p
\end{equation}
into the KdV - Burgers equation  Eq.\eqref{SMM} and finding the
order of the pole $p=2$ for the solution $u(x,t)$ and functions
$u_0(x,t)$ and $u_1(x,t)$ we obtain the transformation
\cite{Kudryashov88}
\begin{equation}
\label{SMM1}u(x,t)=12 \beta \frac{\partial^2 \ln{F}}{\partial x^2}-
\frac{12\,\mu}{5}\,\frac{\partial \ln{F}}{\partial x}+u_2.
\end{equation}

Assuming \cite{Kudryashov88}
\begin{equation}
\label{SMM2} F=1+C_1\,\exp{(k\,x-\omega\,t)}\qquad u_2=C_2,
\end{equation}
(where $C_1$ and $C_2$ are arbitrary constants), we have the system
of algebraic equations with respect to $k$ and $\omega$. Solving
this algebraic system  we have the values of $k$ and $\omega$ as
follows
 \begin{equation}
\label{SMM3}k_{1,2}=\mp\frac{\mu}{5\,\beta},\qquad
\omega_{1,2}=\mp\frac{\mu\,C_2}{5\,\beta}-\frac{6\,\mu^3}{125
\,\beta^2}.
\end{equation}

For these values of $k$ and $\omega$ the solitary wave solutions for
the KdV - Burgers equation takes the form \cite{Kudryashov88}
\begin{equation}\begin{gathered}
\label{SMM4}u(x,t)=C_2-\frac{12\,\mu}{5}\,\frac{\partial }{\partial
x}\ln{(1+C_1\,\exp{ (k\,x-\omega\,t)})}+\\
\\
+12 \,\beta\, \frac{\partial^2 }{
\partial x^2}\ln{(1+C_1\,\exp{(k\,x-\omega\,t)})}.
\end{gathered}\end{equation}
We can see that solution \eqref{SMM4} satisfies the nonlinear
ordinary differential equation in the form
\begin{equation}
\label{SMM4a}\beta\,k^3\,U_{\xi\xi\xi}-\mu\,k^2\,U_{\xi\xi}+
k\,U\,U_{\xi}-\omega\,U_{\xi}=0, \quad \xi=k\,x-\omega\,t.
\end{equation}
We have found exact solution of the KdV --- Burgers equation in
essence using the travelling wave solutions although we tried to
look for more general solutions. Many researches look for the exact
solutions taking directly into account the travelling wave
solutions.

\emph{\textbf{Example 1b.}} \emph{Application of the Riccati
equation as the simplest equation (the first variant)}
\cite{Kudryashov05, Kudryashov05a, Kudryashov07, Kudryashov08,
Kudryashov08b}.

Solution \eqref{SMM4} can be written as
\begin{equation}\begin{gathered}
\label{SMM5}u(\xi)=C_2\,+\frac{C_1\,(60 \beta
\,k^2-12\,\mu\,k)\,\exp{(\xi)}}{5\,(1+C_1\,\exp{(\xi)})} -\frac{ 12
\beta\,C_1^2\,k^2\,\exp{(2\,\xi)}} {(1+C_1\,\exp{(\xi))^2}}.
\end{gathered}\end{equation}
Consider the expression
\begin{equation}\begin{gathered}
\label{SMM6}H=\frac{C_1\,\exp{(\,\xi)}}{1+C_1\,\exp{(\,\xi)}}.
\end{gathered}\end{equation}
Taking the derivative of the function $H$ with respect to $\xi$ we
get
\begin{equation}\begin{gathered}
\label{SMM6a}H_{\xi}=-\,H^2+\,H.
\end{gathered}\end{equation}
We obtain that solution \eqref{SMM5} can be written in the form
\begin{equation}\begin{gathered}
\label{SMM6b}u(\xi)=C_2\,+\,\left(12 \beta
\,k^2-\frac{12\,\mu\,k}{5}\right)\,H -12\, \beta\,k^2\,H^2.
\end{gathered}\end{equation}
Eq.\eqref{SMM6b} means that one can look for the solution of the KdV
- Burgers equation using the expression
\begin{equation}\begin{gathered}
\label{SMM6c}u(\xi)=A_0\,+\,A_1\,H +A_2\,H^2,
\end{gathered}\end{equation}
where $H$ is the solution of Eq.\eqref{SMM6a}. It is seen, that the
application of the simplest equation method with the Riccati
equation gives the same results as the truncated expansion method.

\emph{\textbf{Example 1c.}} \emph{Application of the tanh - function
method} \cite{Lan01, Lou01, Malfliet01, Parkes01}.

Expression \eqref{SMM6} can be transformed as follows
\begin{equation}\begin{gathered}
\label{SMM7a}H=\frac{C_1\,\exp{(\xi)}}{1+C_1\,\exp{(\xi)}}=\frac{1}{2}
\left(\frac{2\,\exp{(\xi-\xi_0)}}
{1+\exp{(\xi-\xi_0)}}-1\right)+\frac{1}{2}=\\
\\
=\frac{1}{2}\left(1+
\tanh{\left(\frac{\xi-\xi_0}{2}\right)}\right),\quad
C_1=\exp{(-\xi_0)}.
\end{gathered}\end{equation}
From \eqref{SMM7a} we have that solution \eqref{SMM5} of the KdV -
Burgers equation can be written in the form
\begin{equation}\begin{gathered}
\label{SMM7b}u(\xi)=C_2-\frac{6\,\mu\,k}{5}+3\beta\,k^2-
\frac{6\,\mu\,k}{5}\tanh{\left(\frac{\xi-\,\xi_0}{2}\right)}-\\
\\
-3\,\beta\,k^2\,\tanh^2{\left(\frac{\xi-\,\xi_0}{2}\right)}.
\end{gathered}\end{equation}

Substituting $k_{1,2}$ and $\omega_{1,2}$ into solution
\eqref{SMM7b} we have the solution of the KdV - Burgers equation in
the form \cite{Kudryashov88}
\begin{equation}\begin{gathered}
\label{SMM7bb}u(\xi)=C_2\pm\frac{6\mu^2}{25\,\beta}\left(1+\tanh{\left(
\frac{\xi-\xi_0}{2}\right)}\right)
+\frac{3\mu^2}{25\beta}\left(1+\tanh^2{\left(\frac{\xi-\,\xi_0}{2}\right)}\right),
\\
\\
\xi=\mp\frac{\mu}{5\,\beta}\,x -
\left(\frac{6\,\mu^3}{125\,\beta^2}\pm
\frac{\mu\,C_2}{5\,\beta}\right)\,t.
\end{gathered}\end{equation}
We obtain that the solution of the KdV - Burgers equation can be
found as the sum of the hyperbolic tangents
\begin{equation}\begin{gathered}
\label{SMM7c}u(\xi)=A_0+A_1\,\tanh{[m\,(\xi-\,\xi_0)]}+A_2\,
\tanh^2{[m\,(\xi-\,\xi_0)]},
\end{gathered}\end{equation}
where $m$ is unknown parameter.

We have obtained that the tanh - function method is equivalent in
essence to the truncated expansion method. As this fact takes place
we can see that the maximum power of the hyperbolic tangent in
\eqref{SMM7c} coincides with the order of the pole for the solution
of the KdV - Burgers equation.

\emph{\textbf{Example 1d.}} \emph{Application of the Riccati
equation as the simplest equation (the second variant).}
\cite{Kudryashov05, Kudryashov05a, Kudryashov07}.

Note that the function
\begin{equation}
\label{SMM8}Y(\xi)=m\,\tanh{[m(\xi-\xi_0)]}
\end{equation}
is the general solution of the Riccati equation
\begin{equation}\begin{gathered}
\label{SMM8a}Y_{\xi}=-Y^2+m^2.
\end{gathered}\end{equation}
From \eqref{SMM7c} we obtain, that we can look for the solution of
the KdV - Burgers equation taking into account the formula
\begin{equation}\begin{gathered}
\label{SMM8aa}u(\xi)=A_0+A_1\,Y+A_2\,Y^2,
\end{gathered}\end{equation}
where $Y$ satisfies Eq.\eqref{SMM8a}. The maximum power of the
function $Y$ in \eqref{SMM8aa} coincides with the pole of the
solution for the KdV - Burgers equation.

\emph{\textbf{Example 1e.}} \emph{Application of the $G_{\xi}/G$
method} \cite{Wang08, Bekir08a, Zhang08}.

Taking into account the transformation
\begin{equation}
\label{SMM9a}Y=\frac{G_{\xi}}{G}
\end{equation}
we reduce Eq.\eqref{SMM8a} to the linear equation
\begin{equation}\begin{gathered}
\label{SMM9b}G_{\xi\xi}-\nu^2\,G=0.
\end{gathered}\end{equation}
As this fact takes place expression \eqref{SMM8aa} can be written in
the form
\begin{equation}\begin{gathered}
\label{SMM9bb}u(\xi)=A_0+A_1\,\frac{G_{\xi}}{G}+A_2\,
\left(\frac{G_{\xi}}{G}\right)^2.
\end{gathered}\end{equation}
This means, that we can search for the exact solutions of the KdV -
Burgers equation in the form \eqref{SMM9bb}, where $G(\xi)$
satisfies Eq.\eqref{SMM9b}. We obtain, that application of the
$G_{\xi}/G$ method is equivalent to application of the simplest
equation method with the Riccati equation, to the tanh - function
method and to the truncated expansion method.

Note that the $G_{\xi}/G$ method follows directly from the truncated
expansion method because solution \eqref{SMM1} can be written as
\begin{equation}\begin{gathered}
\label{SMM9c}u(x,t)=u_2-\frac{12\,\mu}{5}\,
\frac{F_x}{F}-12\,\beta\,\frac{F_x^2}{F^2}+12\,\beta\,\frac{F_{xx}}{F}.
\end{gathered}\end{equation}
Assuming that the function $F(x,t)$ satisfies linear equation of the
second order
\begin{equation}\begin{gathered}
\label{SMM9d}F_{xx}=m_0\,F_x+m_1\,F+m_2,
\end{gathered}\end{equation}
where $m_0$, $m_1$ and $m_2$ are unknown parameters we have the
expression \eqref{SMM9bb} for $F_{\xi}/F$. So, the $G_{\xi}/G$ -
method coincides with the truncated expansion method if we use the
travelling wave solutions.

\emph{\textbf{Example 1f.}} \emph{Application of the tanh --- coth
method} \cite{Wazwaz07, Abdou07, Wazzan09}.

Assuming in \eqref{SMM7c} $\xi_0=\frac{i\,\pi}{2}$,  we obtain the
formula
\begin{equation}
\label{SMM10}u(\xi)=A_0+A_1\,\coth{(\nu\,\xi)}+A_2\,\coth^2{(\nu\,\xi)}.
\end{equation}
This means that we can search for the solution of the KdV - Burgers
equation in the form \eqref{SMM10}.

Using the identity for the hyperbolic functions in the form
\begin{equation}
\label{SMM10a}2\,\coth{(\nu\,\xi)}=\tanh{\left(\frac{\nu\,\xi}{2}\right)}+
\coth{\left(\frac{\nu\,\xi}{2}\right)},
\end{equation}
and substituting \eqref{SMM10a} into \eqref{SMM10} we have the
$\tanh$ - $\coth$ method to look for the solitary wave solutions of
the KdV - Burgers equation
\begin{equation}\begin{gathered}
\label{SMM10b}u(\xi)=B_0+B_1\,\tanh{\left({\nu_1\,\xi}\right)}+B_2
\coth{\left({\nu_1\,\xi}\right)}+\\
\\
+B_3\,\tanh^2{\left({\nu_1\,\xi}\right)}+
B_4\coth^2{\left({\nu_1\,\xi}\right)}.
\end{gathered}\end{equation}
We have obtained, that the tanh - coth method is equivalent to the
truncated expansion method too and we cannot find new exact
solutions of the KdV - Burgers equation by the tanh - coth method.

\emph{\textbf{Example 1g.}} \emph{Application of the Exp - function
method} \cite{He01, He02, Wu01, Ebaid01, Elwakil01}.

Solution \eqref{SMM5} of the KdV - Burgers equation can be written
in the form
\begin{equation}\begin{gathered}
\label{SMM11a}u(\xi)=\frac{5C_2e^{-\xi}+C_1(60\beta k^2-12\mu k+
10C_2)+5C_1^2(C_2-12\beta k^2)e^{\xi}}{5\,e^{-\xi}+
10\,C_1+5\,C_1^2\,e^{\,\xi}}.
\end{gathered}\end{equation}
 This solution can be found if we look for a solution of the KdV --- Burgers
 equation using the Exp - function method in the form
\begin{equation}\begin{gathered}
\label{SMM11b}u(\xi)=\frac{a_0\,e^{-\xi}+a_1+a_2\,e^{\xi}}{b_0\,e^{-\xi}+
b_1+b_2\,e^{\,\xi}}.
\end{gathered}\end{equation}
Using the Exp - function method to search for the exact solutions of
the KdV - Burgers equation one can have again the solitary wave
solutions \eqref{SMM1}  with the functions \eqref{SMM2} and the
parameters \eqref{SMM3}. Studying the application of the Exp -
function method we obtain that this method provides no new exact
solutions of nonlinear differential equations in comparison with
other methods. Moreover the Exp - function method  do not allow us
to find the order of the pole for the solution of nonlinear
differential equation \cite{Kudryashov09a}. Usually the authors use
a few variants of fractions with the sum of exponential functions to
look for the solitary wave solutions.

\section{Second error: some authors do not use the known general solutions of
ordinary differential equations}

Many authors look for the solutions of nonlinear evolution equations
using the travelling waves. As this fact takes place, these authors
obtain nonlinear ordinary differential equations and search for the
solutions of these equations. However nonlinear ordinary
differential equations had been studied very well and the solutions
of them were obtained many years ago. However some of the authors do
not use the well - known general solutions of the ordinary
differential equations. As a result these authors obtain the
solutions that have already been found by other scientists.

The second error can be formulated as follows.

\emph{\textbf{Second error}. Some authors search for solutions of
nonlinear differential equations, but do not use the known general
solutions of these equations.}

\emph{\textbf{Example 2a.}} \emph{Reduction of the (2+1) -
dimensional Burgers equation by Li and Zhang}  \cite{Li}
\begin{equation}
\label{BB1}
u_{t}=u\,u_{y}+\alpha\,v\,u_x+\beta\,u_{yy}+\alpha\,\beta\,u_{xx},
\qquad u_{x}=v_{y}.
\end{equation}

The authors \cite{Li} considered the wave transformations in the
form
\begin{equation}
\label{BB2}u(x,y,t)=U(\xi),\quad v(x,y,t)=V(\xi), \quad
\xi=k(x+ly+\lambda\,t),
\end{equation}
and obtained the system of the ordinary differential equations
\begin{equation}
\label{BB3}\beta\,k\,(\alpha+l^2)\,U_{\xi\xi}+\alpha\,
V\,U_{\xi}+l\,U\,U_{\xi}-\lambda\,U_{\xi}=0,
\end{equation}
\begin{equation}
\label{BB4} U_{\xi}-l\,V_{\xi}=0.
\end{equation}
Li and Zhang \cite{Li} proposed "a generalized multiple Riccati
equation rational expansion method" to construct "a series of exact
complex solutions" for the system of equations \eqref{BB3} and
\eqref{BB4}. The authors  \cite{Li} found "new complex solutions" of
the (2+1) - dimensional Burgers equation and "brought out rich
complex solutions".

However, integrating Eq.\eqref{BB4} with respect to $\xi$ we have
\begin{equation}
\label{BB5} U=l\,V+C_1,
\end{equation}
where $C_1$ is an arbitrary constant. Substituting $U$ into
Eq.\eqref{BB3} we obtain
\begin{equation}\begin{gathered}
\label{BB6}\beta\,k\,(\alpha+l^2)\,V_{\xi\xi}+(\alpha+l^2)\,V\,V_{\xi}+
(C_1\,l-\lambda)\,V_{\xi}=0.
\end{gathered}\end{equation}
Integrating Eq.\eqref{BB6} with respect to $\xi$ we get the Riccati
equation in the form
\begin{equation}\begin{gathered}
\label{BB7}V_{\xi}=-\frac{1}{2\beta\,k}\left(V^2+\frac{2\,C_1\,l-2\,\lambda}
{\alpha+l^2}\,V+2\,\beta\,k\,C_2\right).
\end{gathered}\end{equation}
Eq.\eqref{BB7} can be reduced to the form
\begin{equation}\begin{gathered}
\label{BB8}\frac{dV}{d\xi}=-\frac{1}{2\beta\,k}\,(V-V_{1})(V-V_{2}),
\end{gathered}\end{equation}
where $V_{1}$ and $V_2$ are the roots of the algebraic equation
\begin{equation}\begin{gathered}
\label{BB8a}V^2+\frac{2\,C_1\,l-2\,\lambda}
{\alpha+l^2}\,V+2\,\beta\,k\,C_2=0,
\end{gathered}\end{equation}
that take the form
\begin{equation}\begin{gathered}
\label{BB9}V_{1,2}=\frac{\lambda-C_1\,l\pm
\sqrt{(\lambda-C_1\,l)^2-2\,\beta\,k\,C_2\,(\alpha+l^2)^2}}{\alpha+l^2}.
\end{gathered}\end{equation}

Integrating Eq.\eqref{BB8} with respect to $\xi$, we find the
general solution of Eq.\eqref{BB7} in the form
\begin{equation}\begin{gathered}
\label{BB10}V(\xi)=\frac{V_1-V_2\,\exp{\left(\frac{(V_2-V_1)
(\xi-\xi_0)}{2\,\beta\,k}\right)}}{1-\exp{\left(\frac{(V_2-V_1)
(\xi-\xi_0)}{2\,\beta\,k}\right)}},
\end{gathered}\end{equation}
where $\xi_0$ is an arbitrary constant and $U(\xi)$ is determined by
Eq.\eqref{BB5}.

In the paper \cite{Li} the authors found 24 solitary wave solutions
of the system \eqref{BB3} and \eqref{BB4}, but we can see, that
these solutions are useless for researches.

\emph{\textbf{Example 2b.}} \emph{Reduction of the (3+1) -
dimensional Kadomtsev - Petviashvili equation by Zhang}
\cite{ZhangS}
\begin{equation}
\label{KP} u_{xt}+6\,u_{x}^2+6u\,u_{xx}-u_{xxxx}-u_{yy}-u_{zz}=0.
\end{equation}

This equation was considered  by Zhang \cite{ZhangS}, taking the
travelling wave into account: $u=U(\eta)$,
$\eta=k\,x+l\,y+\,s\,z+\omega\,t$. After reduction Zhang obtained
the nonlinear ordinary differential equation in the form
\begin{equation}\begin{gathered}
\label{KP1}
k\,\omega\,U_{\eta\eta}+6\,k^2\,U_{\eta}^2+6\,k^2\,U\,U_{\eta\eta}-
k^4\,U_{\eta\eta\eta\eta}-l^2\,U_{\eta\eta}-s^2\,U_{\eta\eta}=0.
\end{gathered}\end{equation}
The author \cite{ZhangS} applied the Exp-function method and
obtained the solitary wave solutions of Eq.\eqref{KP1}.

However, denoting $p=\frac{k\,\omega-l^2-s^2}{k^4}$ from Eq.
\eqref{KP1} we have  the nonlinear ordinary differential equation
in the form
\begin{equation}\begin{gathered}
\label{KP2} \,U_{\eta\eta\eta\eta}-\frac{6}{k^2}\,U_{\eta}^2-
\frac{6}{k^2}\,U\,U_{\eta\eta}-p\,U_{\eta\eta} =0.
\end{gathered}\end{equation}

Twice integrating Eq.\eqref{KP2} with respect to $\eta$ we have
\begin{equation}\begin{gathered}
\label{KP3} \,U_{\eta\eta}-\frac{3}{k^2}\,U^2-p\,U -
C_1\,\eta+C_2=0,
\end{gathered}\end{equation}
where $C_1$ and $C_2$ are arbitrary constants. This equation is well
known. Using the transformations for $U$ and $\eta$ by formulae
\begin{equation}\begin{gathered}
\label{KP3a}U =k^2\,\left(\frac{k^2}{C_1}\right)^{-\frac{2}{5}}\,w\,-\,\frac{p\,k^2}{6},\\
\\
\eta=\left(\frac{k^2}{C_1}\right)^{\frac{1}{5}}\,z\,+\,\frac{C_2}{C_1}+
\frac{p^2\,k^2}{12\,C_1},
\end{gathered}\end{equation}
we have the first Painlev\'e equation \cite{Ablowitz, Kudryashov}
\begin{equation}\begin{gathered}
\label{KP3b} w_{zz}=3\,w^2+z.
\end{gathered}\end{equation}

The solutions of Eq. \eqref{KP3b} are the Painlev\'e transcendents.

For the case $C_1=0$ from Eq.\eqref{KP3} we obtain the equation in
the form
\begin{equation}\begin{gathered}
\label{KP3c} \,U_{\eta\eta}-\frac{3}{k^2}\,U^2-p\,U +C_2=0.
\end{gathered}\end{equation}

Multiplying Eq.\eqref{KP3c} on $U_{\eta}$ we have
\begin{equation}\begin{gathered}
\label{KP3d} U_{\eta}^2-\frac{2}{k^2}\,U^3-{p}\,U^2
+2\,C_2\,U+C_3=0,
\end{gathered}\end{equation}
where $C_3$ is an arbitrary constant.

The general solution of Eq. \eqref{KP3d} is found via the
Weierstrass elliptic function \cite{Korteweg, Kudryashov09b}. We can
see, that there is no need to look for the exact solutions of Eq.
\eqref{KP1}. This solution is expressed via the general solution of
well known Eqs. \eqref{KP3b} and \eqref{KP3d}.

\emph{\textbf{Example 2c.}} \emph{Reduction of the Ito equation by
Khani} \cite{Khani}
\begin{gather}
\phi_{xtt}+\phi_{xxxxt}+6\,\phi_{xx}\,\phi_{xt}+
3\,\phi_{x}\,\phi_{xxt}+3\,\phi_{xxx}\,\phi_{t}=0.\label{Ito1}
\end{gather}

Eq. \eqref{Ito1} was studied  by Khani \cite{Khani}. The author
looked for the solutions of Eq. \eqref{Ito1} taking the travelling
wave into account
\begin{gather}
  \phi=\psi(\xi), \qquad \psi=k\,(x-\lambda\,t).\label{Ito2}
  \end{gather}

Substituting \eqref{Ito2}  into \eqref{Ito1} when $\lambda\neq 0$
and $k\neq 0$ we obtain the equation in the form
\begin{gather}
  \lambda\,\psi_{\xi\xi\xi}-k^2\,\psi_{\xi\xi\xi\xi\xi}-6\,k\,\psi_{\xi\xi}\,\psi_{\xi\xi}-
  6\,k\,\psi_{\xi}\,\psi_{\xi\xi\xi}=0.\label{Ito3}
  \end{gather}

Twice integrating Eq. \eqref{Ito3} with respect to $\xi$ we have
\begin{gather}
  \lambda\,\psi_{\xi}-k^2\,\psi_{\xi\xi\xi}-3\,k\,(\psi_{\xi})^2+C_3\,\xi+C_4=0,\label{Ito4}
  \end{gather}
where $C_3$ and $C_4$ are arbitrary constants.

The author \cite{Khani} looked for solution of Eq.\eqref{Ito4} when
$C_3=C_4=0$ using the Exp - function method.

In fact, denoting $\psi_{\xi}=V(\xi)$ in Eq.\eqref{Ito4} we get the
following equation
\begin{gather}
 k^2\,V_{\xi\xi}+3\,k\,V^2 - \lambda\,V-C_3\,\xi-C_4=0.\label{Ito5}
  \end{gather}

Eq. \eqref{Ito5} is equivalent to Eq. \eqref{KP3}. The solutions of
this equation are expressed for $C_3\neq0$ as the first Painlev\'e
transcendents \cite{Ablowitz, Kudryashov} (see the previous
example). For $C_3=0$  the solutions of Eq. \eqref{Ito5} can be
obtained using the Weierstrass elliptic function. So we  need not to
search for the solutions of Eq. \eqref{Ito3} as well.

\emph{\textbf{Example 2d.}} \emph{Reduction of the (3+1) -
dimensional Jimbo - Miva equation by \"{O}zi\c{s} and Aslan}
\cite{Ozis02}
\begin{equation}
\label{JM}
u_{xxxy}+3\,u_{y}\,u_{xx}+3\,u_{x}\,u_{xy}+2\,u_{yt}-3\,u_{xz}=0.
\end{equation}

Using the travelling wave $u(x,y,z,t)=U(\xi)$,
$\xi=k\,x+m\,y+r\,z+w\,t$ Eq. \eqref{JM} can be written as the
nonlinear ordinary differential equation
\begin{equation}\begin{gathered}
\label{JM1}
k^3\,m\,U_{\xi\xi\xi\xi}+6\,k^2\,m\,U_{\xi}\,U_{\xi\xi}+(2\,m\,w-3\,k\,r)\,U_{\xi\xi}=0.
\end{gathered}\end{equation}
The authors \cite{Ozis02} applied the Exp - function method to
Eq.\eqref{JM1} to obtain "the exact and explicit generalized
solitary solutions in more general forms".

But integrating Eq.\eqref{JM1} with respect to $\xi$ we obtain
\begin{equation}\begin{gathered}
\label{JM1a}
k^3\,m\,U_{\xi\xi\xi}+3\,k^2\,m\,U_{\xi}^2+(2\,m\,w-3\,k\,r)\,U_{\xi}=C_5,
\end{gathered}\end{equation}
where $C_5$ is an arbitrary constant. Denoting $U_{\xi}=V$ we get
\begin{equation}\begin{gathered}
\label{JM1b}
k^3\,m\,V_{\xi\xi}+3\,k^2\,m\,V^2+(2\,m\,w-3\,k\,r)\,V=C_5.
\end{gathered}\end{equation}
Multiplying Eq.\eqref{JM1b} on $V_{\xi}$ we have the equation in the
form
\begin{equation}\begin{gathered}
\label{JM1c}
V_{\xi}^2+\frac{2}{k}\,V^3+\frac{2\,m\,w-3\,k\,r}{m\,k^3}\,V^2-\frac{2\,C_5}{mk^3}\,V-C_6=0,
\end{gathered}\end{equation}
where $C_6$ is an arbitrary constant. The general solution can be
found using the Weierstrass elliptic function. The solution of
Eq.\eqref{JM1} is found by the integral
\begin{equation}\begin{gathered}
\label{JM1d} U=\int V d \xi.
\end{gathered}\end{equation}
We can see that Eq.\eqref{JM1a} has the general solution
\eqref{JM1d} and all partial cases can be found from the general
solution of Eq.\eqref{JM1d}.

\emph{\textbf{Example 2e.}} \emph{Reduction of the Benjamin - Bona -
Mahony equation} \cite{Ganji}
\begin{equation}\begin{gathered}
u_{t}-u_{xxt}+u_{x}+\left(\frac{u^2}{2}\right)_x=0. \label{BBMB}
\end{gathered}\end{equation}

Eq. \eqref{BBMB} was considered by Ganji and co - authors
\cite{Ganji}. In terms of the travelling wave $u(x,t)=U(\eta)$,
$\eta=k\,x-\omega\,t$ they obtained the equation
\begin{equation}\begin{gathered}
k^2\,\omega\,\,U_{\eta\eta\eta}+k\,U\,U_{\eta}+(k-\omega)\,U_{\eta}=0.
\label{BBMB1a}
\end{gathered}\end{equation}
Using the Exp - function method the authors \cite{Ganji} looked for
the solitary wave solutions of Eq.\eqref{BBMB1a}.

Integrating Eq. \eqref{BBMB1a} with respect to $\eta$ we have
\begin{equation}\begin{gathered}
k^2\,\omega\,\,U_{\eta\eta}+\frac12\,k\,U^2+(k-\omega)\,U+C_1=0.
\label{BBMB2a}
\end{gathered}\end{equation}
Multiplying Eq.\eqref{BBMB2a} on $U_{\eta}$ and integrating the
result with respect to $\eta$ again we obtain the equation
\begin{equation}\begin{gathered}
k^2\,\omega\,\,U_{\eta}^2+\frac{k}{3}\,U^3+(k-\omega)\,U^2+2\,C_1+C_2=0.
  \label{BBMB3a}
  \end{gathered}\end{equation}

The solution of Eq.\eqref{BBMB3a} can be given by the Weierstrass
elliptic function \cite{Ablowitz, Kudryashov, Korteweg,
Kudryashov09b}. We can see that the authors \cite{Ganji} obtained
the known solitary wave solutions of Eq.\eqref{BBMB}.

\emph{\textbf{Example 2f.}} \emph{Reduction of the Sharma - Tasso -
Olver equation by Erbas and Yusufoglu}\cite{Erbas01}
\begin{equation}\begin{gathered}
u_{t}+\alpha\,\left(u^3\right)_x+\frac32\,\alpha\,\left(u^2\right)_{xx}+\alpha\,u_{xxx}=0.
\label{STO}
\end{gathered}\end{equation}

Taking the travelling wave $u(x,t)=U(\xi)$, $\xi=\mu\,(x-c\,t)$ the
authors \cite{Erbas01} obtained the reduction of Eq.\eqref{STO} in
the form
\begin{equation}\begin{gathered}
\alpha\,\mu^2\,U_{\xi\xi}+3\,\alpha\,\mu\,U\,U_{\xi}+\alpha\,U^3-c\,U=0.
\label{STO1}
\end{gathered}\end{equation}
The authors of \cite{Erbas01} used the Exp - function method to find
"new solitonary solutions", but they left out of their account that
using the transformation
\begin{equation}\begin{gathered}
U=\mu\,\frac{F_{\xi}}{F} \label{STO2}
\end{gathered}\end{equation}
Eq.\eqref{STO1} can be transformed to the linear equation
\begin{equation}\begin{gathered}
F_{\xi\xi\xi}-\frac{c}{\alpha\,\mu^2}\,F_{\xi}=0. \label{STO3}
\end{gathered}\end{equation}
The solution of Eq.\eqref{STO3} takes the form
\begin{equation}\begin{gathered}
F{(\xi)}=C_1+C_2\,\exp{\left\{\frac{\xi\,\sqrt{c}}{\mu\,\sqrt{\alpha}}\right\}}+
C_3\,\exp{\left\{-\frac{\xi\,\sqrt{c}}{\mu\,\sqrt{\alpha}}\right\}}.
\label{STO4}
\end{gathered}\end{equation}
Substituting the solution \eqref{STO4} into the transformation
\eqref{STO2} we obtain the solution of Eq.\eqref{STO1} in the form
\begin{equation}\begin{gathered}
u{(\xi)}=\frac{\sqrt{c}}{\sqrt{\alpha}}\,\,\frac{C_2\,
\exp{\left\{\frac{\xi\,\sqrt{c}}{\mu\,\sqrt{\alpha}}\right\}}-
C_3\,\exp{\left\{-\frac{\xi\,\sqrt{c}}{\mu\,\sqrt{\alpha}}\right\}}}
{C_1+C_2\,\exp{\left\{\frac{\xi\,\sqrt{c}}{\mu\,\sqrt{\alpha}}\right\}}+
C_3\,\exp{\left\{-\frac{\xi\,\sqrt{c}}{\mu\,\sqrt{\alpha}}\right\}}}
\label{STO5}.
\end{gathered}\end{equation}

 Certainly all the solutions obtained by means of the Exp
- function method can be found from solution \eqref{STO5}.

\emph{\textbf{Example 2g.}} \emph{Reduction of the dispersive long
wave equations by Abdou} \cite{Abdou07}
\begin{equation}\begin{gathered}
\label{LWE} v_{t}+v\,v_x+w_{x}=0, \\
\\
w_t+(vw)_x+\frac13\,v_{xxx}=0.
\end{gathered}\end{equation}

Using the wave transformations $w(x,t)=\sigma(\xi)$,
$v(x,t)=\phi(\xi)$, $\xi=k(x+\lambda\,t)$, the system of equations
\eqref{LWE} can be written in the form
\begin{equation}\begin{gathered}
\label{LWE1} \lambda\,\phi_{\xi}+\phi\,\phi_{\xi}+\sigma_{\xi}=0,
\end{gathered}\end{equation}
\begin{equation}\begin{gathered}
\label{LWE2}
\lambda\,\sigma_{\xi}+(\sigma\,\phi)_{\xi}+\frac{k^2}{2}\phi_{\xi\xi\xi}=0.
\end{gathered}\end{equation}

Abdou \cite{Abdou07} looked for solutions of the system of equations
\eqref{LWE1} and \eqref{LWE2} taking "the extended tanh method" into
account.

Integrating Eqs. \eqref{LWE1} and \eqref{LWE2} with respect to $\xi$
we obtain

\begin{equation}\begin{gathered}
\label{LWE3} \sigma=-C_1-\lambda\,\phi-\frac12\,\phi^2,
\end{gathered}\end{equation}
\begin{equation}\begin{gathered}
\label{LWE4}
\lambda\,\sigma+(\sigma\,\phi)+\frac{k^2}{2}\phi_{\xi\xi}+C_2=0.
\end{gathered}\end{equation}

Substituting the value of $\sigma$ from \eqref{LWE3} into
Eq.\eqref{LWE4} we have
\begin{equation}\begin{gathered}
\label{LWE5}\phi_{\xi\xi}-\frac{1}{k^3}\phi^3-\frac{3\,\lambda}{k^3}\phi^2-
\frac{2(C_1+\lambda)}{k^3}\phi+\frac{2\,C_2}{k^3}=0.
\end{gathered}\end{equation}

Multiplying Eq.\eqref{LWE5} on $\phi_{\xi}$ and integrating the
equation with respect to $\xi$, we have the equation in the form

\begin{equation}\begin{gathered}
\label{LWE6}\phi_{\xi}^2-\frac{1}{2\,k^3}\phi^4-\frac{3\,\lambda}{k^3}\phi^3-
\frac{(C_1+\lambda)}{k^3}\phi^2+\frac{2\,C_2}{k^3}\,\phi+C_3=0.
\end{gathered}\end{equation}

The general solution of Eq.\eqref{LWE6} is determined via the Jacobi
elliptic function \cite{Kudryashov}. The variable $\sigma$ is found
by the formula \eqref{LWE3} and there is no need to look for the
exact solutions of the Eqs.\eqref{LWE1} and \eqref{LWE2}.

Unfortunately we have many similar examples. Some of them are also
presented in our recent work \cite{Kudryashov09b}.

\section{Third error: some authors omit arbitrary constants after integration of equation}

Reductions of nonlinear evolution equations to nonlinear ordinary
differential equations can be often integrated. However, some
authors assume, that the arbitrary constants of integration are
equal to zero. This error potentially leads to the loss of the
arbitrary constants in the final expression. So the solution
obtained in such way is less general than it could be. The third
common error can be formulated as follows.

\emph{\textbf{Third error}. Some authors omit the arbitrary
constants after integrating of the nonlinear ordinary differential
equations.}

\emph{\textbf{Example 3a.}} \emph{Reduction of the Burgers equation
by Soliman}\cite{Soliman01}.

Soliman \cite{Soliman01} considered the Burgers equation
\begin{equation}
\label{B} u_t+u\,u_x-\nu\, u_{xx}=0
\end{equation}
to solve this equation by so called "the modified extended tanh -
function method".

It is well known, that by using the Cole-Hopf transformation
\cite{Cole01, Hopf01}
\begin{equation}\begin{gathered}
\label{B.01}u=-2\,\nu\,\frac{\partial}{\partial x} \ln{F},
\end{gathered}\end{equation}
we can write the equality
\begin{equation}\begin{gathered}
\label{B.02}u_t+u\,u_x-\nu\,u_{xx}=-2\,\nu\,\frac{\partial}{\partial
x}\left(\frac{F_t-\nu\,F_{xx}}{F}\right).
\end{gathered}\end{equation}
From the last relation we can see, that each solution of the heat
equation
\begin{equation}\begin{gathered}
\label{B.04}F_t-\nu\,F_{xx}=0,
\end{gathered}\end{equation}
gives the solution of the Burgers equation by formula \eqref{B.01}.

However  to find the solutions of the Burgers equation Soliman
\cite{Soliman01} used the travelling wave solutions $u(x,t)=U(\xi)$,
$\xi=x-c\,t$ and from Eq.\eqref{B} after integration with respect to
$\xi$ the author obtained
 the equation in the form
\begin{equation}\begin{gathered}
\label{B.07}\nu \,U_{\xi}-\frac{1}{2}\,U^2+c\,U=0.
\end{gathered}\end{equation}
The constant of integration he took to be equal to zero. The general
solution of Eq.\eqref{B.07} takes the form
\begin{equation}\begin{gathered}
\label{B.08}U{(\xi)}=\frac{2\,c\,C_2\,
\exp{\left\{-\,{\frac{c\,\xi}{\nu}}\right\}}}{1+C_2\,
\exp{\left\{-\,{\frac{c\,\xi}{\nu}}\right\}}},
\end{gathered}\end{equation}
where $C_2$ is an arbitrary constant.

The general solution of Eq.\eqref{B.07} has the only arbitrary
constant. But if we take nonzero constant of integration in
Eq.\eqref{B.07}, we can have two arbitrary constants in the
solution.

\emph{\textbf{Example 3b. }} \emph{Reduction  of the (2+1) -
dimensional Konopelchenko - Dubrovsky equation by
Abdou}\cite{Abdou02}
\begin{gather}
u_{t}-u_{xxx}-6\,b\,u\,u_{x}+\frac{3\,a^2}{2}\,u^2\,u_x-3\,v_y+3\,a\,v\,u_x=0,
\qquad u_y=v_x. \label{KD}
\end{gather}
Using the wave solutions
\begin{gather}
u(x,t)=U(\eta), \qquad \eta=k\,x+l\,y+\omega\,t\label{KD1}
  \end{gather}
the author \cite{Abdou02} looked for the exact solutions of
Eq.\eqref{KD}. After integration with respect to $\eta$ he obtained
the second order differential equation
\begin{gather}
\left(\omega-\frac{3\,l^2}{k}\right)\,U-k^3\,U_{\eta\eta}+\left(\frac{3al}{2}-
3\,b\,k\right)\,U^2+\frac{a^2}{2}\,k\, U^3=0,\label{KD2}
\end{gather}
but the zero constant of integration was taken. Abdou \cite{Abdou02}
used the Exp - function method to look for solitary wave solutions
of Eq.\eqref{KD2}.

However multiplying Eq.\eqref{KD2} on $U_{\eta}$ and integrating
this equation with respect to $\eta$ again, we have the equation
\begin{gather}
\,\left(\omega-\frac{3\,l^2}{k}\right)\,U^2-\,k^3\,U_{\eta}^2+\left({al}{}-
2\,b\,k\right)\,U^3+\,\frac{a^2}{4}\,k\, U^4=C_2,\label{KD2a}
\end{gather}
where $C_2$ is a constant of integration. The general solution of
Eq.\eqref{KD2a} is expressed via the Jacobi elliptic function
\cite{Kudryashov}.

\emph{\textbf{Example 3c.}} \emph{Reduction  of the Ito equation by
Wazwaz} \cite{WazwazAM}
\begin{gather}
  v_{xtt}+v_{xxxxt}+6\,v_{xx}\,v_{xt}+
  3\,v_{x}\,v_{xxt}+3\,v_{xxx}\,v_{t}=0\label{WW1}.
  \end{gather}

The author \cite{WazwazAM} looked for the solutions of Eq.
\eqref{WW1} taking into account the travelling wave
\begin{gather}
  v=v(\xi), \qquad \xi=k\,(x-\lambda\,t).\label{WW2}
  \end{gather}

Substituting \eqref{WW2} into \eqref{WW1} Wazwaz obtained when
$\lambda\neq 0$ and $k\neq 0$ the equation in the form
\begin{gather}
  \lambda\,v_{\xi\xi\xi}-k^2\,v_{\xi\xi\xi\xi\xi}-6\,k\,v_{\xi\xi}\,v_{\xi\xi}-
  6\,k\,v_{\xi}\,v_{\xi\xi\xi}=0.\label{WW3}
  \end{gather}

Integrating Eq.\eqref{WW3} twice with respect to $\xi$ one can have
the equation
\begin{gather}
  \lambda\,v_{\xi}-k^2\,v_{\xi\xi\xi}-3\,k\,(v_{\xi})^2+C_8\,\xi+C_9=0,\label{WW4}
  \end{gather}
where $C_8$ and $C_9$ are arbitrary constants. Denoting
$v_{\xi}=V(\xi)$ in Eq. \eqref{WW4} we get the equation
\begin{gather}
 k^2\,V_{\xi\xi}+3\,k\,V^2 - \lambda\,V-C_8\,\xi-C_9=0.\label{WW5}
  \end{gather}

The general solution of this equation was discussed above in example
2b.

However the author \cite{WazwazAM} looked for solutions of Eq.
\eqref{WW4} for $C_8=0$ and $C_9=0$ taking into consideration the
tanh - coth method and did not present the general solution of Eq.
\eqref{WW3}.

\emph{\textbf{Example 3d.}} \emph{Reduction  of the Boussinesq
equation by Bekir} \cite{Bekir08a}
\begin{gather}
  u_{tt}-u_{xx}-(u^2)_{xx}+u_{xxxx}=0 \label{Bu1}.
  \end{gather}
Using the wave variable $\xi=x-c\,t$ the author \cite{Bekir08a} got
the equation
\begin{gather}
u_{\xi\xi}-u^2+(c^2-1)u=0 \label{Bu2}.
  \end{gather}
To look for the solitary wave solutions of Eq.\eqref{Bu2} the author
\cite{Bekir08a} used the $G^{'}/G $ - method, but he have been
omitted two arbitrary constants after the integration.

In fact, from Eq.\eqref{Bu1} we obtain the second order differential
equation  in the form
\begin{gather}
u_{\xi\xi}-u^2+ (c^2-1)u+C_1\,\xi+C_2=0. \label{Bu3}
  \end{gather}
The general solution of Eq.\eqref{Bu3} is expressed at $C_1\neq0$
via the first Painlev\'e transcendents. In the case $C_1=0$
solutions of Eq.\eqref{Bu3} is determined by the Weierstrass
elliptic function. All possible solutions of Eq. \eqref{Bu1} were
obtained in work \cite{Clarkson}.

\section{Fourth error: using some functions in finding exact
solutions some authors lose arbitrary constants}

Some authors do not include the arbitrary constants in finding the
exact solutions of nonlinear differential equations. As a result
these authors obtain many solutions, that can be determined as the
only solution using some arbitrary constants. The arbitrary
constants can be included, if we use the general solutions of the
known differential equations. Sometimes this error can be corrected,
using the property of the autonomous differential equation. So, the
fourth error can be formulated as follows.

\emph{\textbf{Fourth error}. Using some functions in finding the
exact solutions  of nonlinear differential equations some authors
lose the arbitrary constants}.

Let us explain this error. Consider an ordinary differential
equation in the general form
\begin{equation}
\label{A} E(Y,Y_{\xi},...) =0.
\end{equation}

Let us assume that Eq. \eqref{A} is autonomous and this equation
admits the shift of the independent variable $\xi \rightarrow
\xi+C_2$ (where $C_2$ is an arbitrary constant). This means that
 the constant $C_2$ added to the variable $\xi$ in Eq.\eqref{A} does
not change the form of this equation. In this case  Eq.\eqref{A} is
invariant under the shift of the independent variable.

Taking this property into account, we obtain the advantage for the
solution of Eq. \eqref{A}. If we know a solution $Y=f(\xi)$ of Eq.
\eqref{A}, then for the autonomous equation we have a solution
$Y=f(\xi-\xi_0)$ of this equation with additional arbitrary constant
$\xi_0$.

The main feature of the autonomous equation is that the fact that
the solution $Y=f(\xi-\xi_0)$ is more general, then $Y=f(\xi)$.

The error discussed often leads to a huge amount of different
expressions for the solutions of nonlinear differential equations
instead of choosing one solution with an arbitrary constant $\xi_0$.

The application of the tanh - function method \cite{Lou01,
Malfliet01, Parkes01} for finding the exact solutions allows us to
have the special solutions of nonlinear differential equations as a
sum of hyperbolic tangents $\tanh{(\xi)}$. However for the
autonomous equation such types of the solutions can be taken as the
more general solution in the form $\tanh{(\xi-\xi_0)}$.

\emph{\textbf{Example 4a. }}\emph{Solution of the Riccati equation}
\begin{equation}
\label{R} Y_{\xi}=\beta(1+Y^2).
\end{equation}

Eq. \eqref{R} is of the first order, therefore the general solution
of Eq. \eqref{R} depends on the only arbitrary constant. We can meet
a lot of "different" solutions of Eq. \eqref{R}. For example
\begin{equation}\begin{gathered}
\label{S1} Y_1(\xi)=\tan {\{\beta\,\xi\}},
\end{gathered}\end{equation}
\begin{equation}\begin{gathered}
\label{S2} Y_2(\xi)=- \cot{\{\beta\,\xi\}},
\end{gathered}\end{equation}
\begin{equation}\begin{gathered}
\label{S3} Y_3(\xi)=\tan{\{2\beta \xi\}}-\sec{\{2\beta \xi\}},
\end{gathered}\end{equation}
\begin{equation}\begin{gathered}
\label{S4} Y_4(\xi)=\csc{\{2\beta \xi\}}-\cot{\{2\beta \xi\}}.
\end{gathered}\end{equation}

In fact, the general solution of Eq. \eqref{R}
takes the form
\begin{equation}
\label{S} Y(\xi)=\tan{\{{\beta}(\xi-\xi_0)\}}.
\end{equation}

All solutions \eqref{S1} - \eqref{S4} can be obtained from the
general solution \eqref{S} of the Riccati equation, because of the
following equalities

\begin{equation}\begin{gathered}
\label{Relations}-\cot\beta x=\tan(\beta x+\pi/2),
\\
\tan 2\beta x-\sec 2\beta x=\tan(\beta x-\pi/4),
\\
\csc 2\beta x-\cot 2\beta x=\tan{\beta x}.
\end{gathered}\end{equation}

\emph{\textbf{Example 4b. }} \emph{Solution of the Cahn - Hilliard
equation by Ugurlu and Kaya} \cite{Ugurlu}
\begin{equation}
\label{CH} u_t+u_{xxxx}=\left(u^3-u\right)_{xx}+\,u_x.
\end{equation}

Eq.\eqref{CH} was considered in \cite{Ugurlu} by means of the
modified extended tanh - function method by Ugurlu and Kaya. Using
the travelling wave
\begin{equation}
\label{CH1} u(x,t)=u(z), \qquad z=x+c\,t,
\end{equation}
the authors obtained the exact solutions of the equation
\begin{equation}
\label{CH2}
c\,u_z+u_{zzzz}-6\,u\,\left(u_z\right)^2-3\,u^2\,u_{zz}+u_{zz}-u_z=0.
\end{equation}
They found eight solutions of Eq.\eqref{CH2} at $c=1$. Six solutions
are the following
\begin{equation}
\label{CH3} u_1=\coth{\left\{\frac{\sqrt{2}}{2}z\right\}},\quad
u_2=-\coth{\left\{\frac{\sqrt{2}}{2}z\right\}},\quad
u_3=\tanh{\left\{\frac{\sqrt{2}}{2}z\right\}},
\end{equation}
\begin{equation}
\label{CH4} \quad
u_4=\frac12\,\left(\tanh{\left\{\frac{z}{2\sqrt{2}}\right\}}+
\coth{\left\{\frac{z}{2\sqrt{2}}\right\}}\right),\quad
u_5=-\tanh{\left\{\frac{\sqrt{2}}{2}z\right\}},
\end{equation}

\begin{equation}
\label{CH5}u_6=-\frac12\,\left(\tanh{\left\{\frac{z}{2\sqrt{2}}\right\}}+
\coth{\left\{\frac{z}{2\sqrt{2}}\right\}}\right).
\end{equation}

However all these solutions can be written as the only solution with
an arbitrary constant $z_0$
\begin{equation}
\label{CH8} u=\tanh{\left\{\frac{\sqrt{2}}{2}(z-z_0)\right\}}.
\end{equation}
Note, that Eq.\eqref{CH2} at $c=1$ takes the form
\begin{equation}
\label{CH2a}
u_{zzzz}-6\,u\,\left(u_z\right)^2-3\,u^2\,u_{zz}+u_{zz}=0.
\end{equation}
Twice integrating Eq.\eqref{CH2a} with respect to $z$, we have
\begin{equation}
\label{CH2b} u_{zz}-2\,u^3+u+C_1\,z+C_2=0,
\end{equation}
where $C_1$ and $C_2$ are arbitrary constants. At $C_1=0$ the
solution of Eq.\eqref{CH2b} is expressed via the Jacobi elliptic
function.

\emph{\textbf{Example 4c. }} \emph{Solution of the KdV - Burgers
equation by Soliman} \cite{Soliman01}
\begin{equation}
\label{KdVB} u_t+\varepsilon\,u\,u_x-\nu\,u_{xx}+\mu\,u_{xxx}=0.
\end{equation}

Using the travelling wave and "the modified extended tanh - function
method" the author \cite{Soliman01} obtained four solutions of
Eq.\eqref{KdVB}. Three of them take the form
\begin{equation}
\label{KdVB1} u(x,t)=\frac{\nu^2}{25\varepsilon
\mu}\left[9-6\coth{\left(\frac{\nu\xi}{10\mu}\right)}-
3\coth^2{\left(\frac{\nu\xi}{10\,\mu}\right)}\right],\quad \xi=x-
\frac{6\nu^2t}{25\mu};
\end{equation}

\begin{equation}
\label{KdVB2} u(x,t)=\frac{\nu^2}{25\varepsilon
\mu}\left[9-6\tanh{\left(\frac{\nu\xi}{10\mu}\right)}-
3\tanh^2{\left(\frac{\nu\xi}{10\mu}\right)}\right],\quad \xi=x-
\frac{6\nu^2t}{25\mu};
\end{equation}

\begin{equation}\begin{gathered}
\label{KdVB3} u(x,t)=\frac{3\nu^2}{25\varepsilon
\mu}\left[1-\frac{4}{10}\left(\tanh{\left(\frac{\nu\xi}{20\mu}\right)}+
\coth{\left(\frac{\nu\xi}{20\,\mu}\right)}\right)\right. -\\
\\
-\left.\frac{1}{10}
\left(\tanh^2{\left(\frac{\nu\xi}{20\mu}\right)}+
\coth^2{\left(\frac{\nu\xi}{20\,\mu}\right)}\right)\right],\quad
\xi=x- \frac{6\nu^2t}{25\mu}. \end{gathered}\end{equation}

All these solutions can be written in the form
\begin{equation}\begin{gathered}
\label{KdVB4} u(x,t)=\frac{3\,\nu^2}{25\varepsilon
\mu}\left[3-2\tanh{\left(\frac{\nu\xi}{10\mu}-\xi_0\right)}-
\tanh^2{\left(\frac{\nu\,\xi}{10\,\mu}-\xi_0\right)}\right],\\
\\
\xi=x- \frac{6\nu^2t}{25\mu}.
\end{gathered}\end{equation}

Assuming $\xi_0=0$ in \eqref{KdVB4}, we obtain solution
\eqref{KdVB2}. In the case $\xi_0=\frac{i\pi}{2}$ in \eqref{KdVB4}
we have solution \eqref{KdVB1}. Taking into account the formulae
\begin{equation}\begin{gathered}
\label{KdVB5} \tanh{(k\xi)}+\coth{(k\xi)}=2\,\coth{(2\,k\,\xi)},
\end{gathered}\end{equation}
\begin{equation}\begin{gathered}
\label{KdVB6}
\tanh^2{(k\xi)}+\coth^2{(k\xi)}=4\,\coth^2{(2\,k\,\xi)}-2,
\end{gathered}\end{equation}
we can transform solution \eqref{KdVB4} into solution \eqref{KdVB3}.

We can see, that these solutions do not differ, if we take the
constant $\xi_0$ into account in one of these solutions.

The fourth solution by Soliman \cite{Soliman01} can be simplified as
well. All the solutions by the KdV --- Burgers equation by Soliman
coincides with \eqref{SMM7bb}.

\emph{\textbf{Example 4d. }} \emph{Solution of the combined KdV -
mKdV equation by Bekir} \cite{Bekir09}
\begin{equation}
\label{KdVmK} u_t+p \,u\,u_x+q\,u^2\,u_{x}+r\,u_{xxx}=0.
\end{equation}

Using the extended tanh method the author \cite{Bekir09} have
obtained the following solutions of Eq.\eqref{KdVmK}
\begin{equation}
\label{KdVmK1} u=\frac{12 r}{p}[1+\tanh{(x-4rt)}],
\end{equation}
\begin{equation}
\label{KdVmK2} u=\frac{12 r}{p}[1+\coth{(x-4rt)}],
\end{equation}
\begin{equation}
\label{KdVmK3} u=\frac{24
r}{p}[(21+\tanh{(x-4rt)})+(1+\coth{(x-4rt)})].
\end{equation}

All these solutions can be written as the only solution with an
arbitrary constant
\begin{equation}
\label{KdVmK1} u=\frac{12 r}{p}[1+\tanh{(x-4rt+\varphi_0)}].
\end{equation}

\emph{\textbf{Example 4e. }} \emph{Solution of the coupled Hirota
--- Satsuma --- KdV equation by Bekir} \cite{Bekir08}
\begin{equation}
\label{HSK} u_t=\frac14\,u_{xxx}+3\,u\,u_x-6\,v\,u_x,
\end{equation}
\begin{equation}
\label{HSK1} v_t=-\frac12\,v_{xxx}-3\,u\,v_x.
\end{equation}
The system of Eqs.\eqref{HSK} and \eqref{HSK1} was studied by means
of the tanh - coth method by Bekir \cite{Bekir08}. Using the
travelling wave $ \xi=\,(x-\beta\,t)$ the author obtained the system
of equations
\begin{equation}\begin{gathered}
\label{HSK2}\frac14\,U_{\xi\xi\xi}+3\,U\,U_{\xi}-6\,V\,V_{\xi}+\beta\,U_{\xi}=0\\
 \frac12\,V_{\xi\xi\xi}+3\,U\,V_{\xi}-\beta\,V_{\xi}=0,
\end{gathered}\end{equation}
and gave ten solitary wave solutions. However all the solitary wave
solutions of the system of equations \eqref{HSK} and \eqref{HSK1} by
Bekir \cite{Bekir08} can be expressed by the formulae
\begin{equation}\begin{gathered}
\label{HSK3} U^{(1)}=\frac{\beta+4\,k^2}{3}-2\,k^2\,\tanh^2{(k\,\xi-\xi_0)},\\
\\
V^{(1)}=\frac{2\,\beta+2\,k^2}{3}-k^2\,\tanh^2{(k\,\xi-\xi_0)},
\quad \xi=x-\beta\,t,
\end{gathered}\end{equation}
(where $k$,  $\xi_0$ and $\beta$ are arbitrary constants)
\begin{equation}\begin{gathered}
\label{HSK4} U^{(2)}=k^2-2\,k^2\,\tanh^2{(k\,\xi-\xi_0)},\\
\\
V^{(2)}=-k^2\,\tanh^2{(k\,\xi-\xi_0)}, \quad \xi=x+k^2\,t, \quad
\beta=-k^2,
\end{gathered}\end{equation}
\begin{equation}\begin{gathered}
\label{HSK5} U^{(3,4)}=\frac{\beta+k^2}{3}-k^2-k^2\,\tanh^2{(k\,\xi-\xi_0)},\\
\\
V^{(3,4)}=\pm\sqrt{\frac{k^2-2\,\beta\,k}{3}}\,\tanh{(k\,\xi-\xi_0)},
\quad \xi=x-\beta\,t.
\end{gathered}\end{equation}

\emph{\textbf{Example 4f. }} \emph{"Twenty seven solution" of the
"generalized Riccati equation" by Xie, Zhang and L\"{u}}
\cite{Xie05}.

The authors \cite{Xie05} "firstly extend" the Riccati equation to
the "general form"
\begin{equation}\begin{gathered}
\label{GRE1}\phi_{\xi}=r+p\,\phi+q\,\phi^2,
\end{gathered}\end{equation}
where $r$, $p$ and $q$ are the parameters. They "fortunately find
twenty seven solutions" of Eq.\eqref{GRE1}.

It was very surprised that the authors are not aware that the
solution of Eq.\eqref{GRE1} was known more then one century ago. It
is very strange but these 27 solutions was repeated by Zhang
\cite{Zhang07a} as the important advantage.

Let us present the general solution of Eq.\eqref{GRE1}. Substituting
\begin{equation}\begin{gathered}
\label{GRE2}\phi=\frac{1}{q}\,V-\frac{p}{2\,q}
\end{gathered}\end{equation}
into Eq.\eqref{GRE1} we have
\begin{equation}\begin{gathered}
\label{GRE3}V_{\xi}=V^2+\frac{4rq-p^2}{4}.
\end{gathered}\end{equation}
Assuming
\begin{equation}\begin{gathered}
\label{GRE3a}V=\frac{\psi_{\xi}}{\psi}
\end{gathered}\end{equation}
in Eq.\eqref{GRE3} we obtain the linear equation of the second order
\begin{equation}\begin{gathered}
\label{GRE4}\psi_{\xi\xi}-\frac{4\,r\,q-p^2}{4}\,\psi=0.
\end{gathered}\end{equation}
The general solution of Eq.\eqref{GRE4} is well known
\begin{equation}\begin{gathered}
\label{GRE5}\psi(\xi)=C_1\,e^{\frac{\xi}{2}\sqrt{4rq-p^2}}+C_2\,
e^{-\frac{\xi}{2}\sqrt{4rq-p^2}}.
\end{gathered}\end{equation}
Using formula \eqref{GRE3a} and \eqref{GRE2} we obtain the solution
of Eq.\eqref{GRE1} in the form
\begin{equation}\begin{gathered}
\label{GRE6}\phi(\xi)=\frac{\sqrt{4rq-p^2}}{2q}\,\,\frac{C_1\,
e^{\frac{\xi}{2}\sqrt{4rq-p^2}}-C_2\,
e^{-\frac{\xi}{2}\sqrt{4rq-p^2}}}{C_1\,e^{\frac{\xi}{2}\sqrt{4rq-p^2}}+C_2\,
e^{-\frac{\xi}{2}\sqrt{4rq-p^2}}}-\frac{p}{2q}.
\end{gathered}\end{equation}

The general solution of Eq.\eqref{GRE1} is found from \eqref{GRE6}.
All 27 solutions by Xie, Zhang and L\"{u} are found from solution
\eqref{GRE6} and  we cannot obtain other solutions.

\emph{\textbf{Example 4g. }} \emph{"New solutions and kinks
solutions" of the Sharma --- Tasso --- Olver equation by Wazwaz}
\cite{Wazwaz07b}.

Using the extended tanh method Wazwaz \cite{Wazwaz07b} have found 18
solitary wave and kink solutons of the Sharma --- Tasso ---- Olver
equation
\begin{equation}\begin{gathered}
\label{OO1}u_t+\alpha\,(u^3)_x+\frac32\,\alpha\,(u^2)_{xx}+\,\alpha\,u_{xxx}=0.
\end{gathered}\end{equation}
Taking the travelling wave solution $u(x,t)=u(\xi)$, $\xi=x-c\,t$
into account the author considered the nonlinear ordinary
differential equation in the form
\begin{equation}\begin{gathered}
\label{OO2}\alpha\,u_{\xi\xi}+3\,\alpha\,u\,u_{\xi}+\alpha\,u^3-c\,u=0.
\end{gathered}\end{equation}
However Eq.\eqref{OO2} can be transformed to the second - order
linear differential equation (see, example 2f)
\begin{equation}\begin{gathered}
\label{OO4}F_{\xi\xi\xi}-\frac{c}{\alpha}\,F_{\xi}=0
\end{gathered}\end{equation}
by the transformation
\begin{equation}\begin{gathered}
\label{OO3}\,u{(\xi)}=\frac{F_{\xi}}{F}.
\end{gathered}\end{equation}
The solution of Eq.\eqref{OO2} is given by formula \eqref{STO5} at
$\mu=1$. Certainly all "new solutions" of the Sharma --- Tasso ---
Olver equation by Wazwaz \cite{Wazwaz07b} are found from
\eqref{STO5} at $\mu=1$.

We can see, that there is no need to write a list of all possible
expressions for the solutions at the given $\xi_0$. It is enough to
present the solution of the equation with an arbitrary constant.
Moreover, the solution with arbitrary constants looks better.

The simple and powerful tool to remove this error is to plot the
graphs of the expressions obtained. The expressions having the same
graphs usually are equivalent.

\section{Fifth error: some authors do not simplify the solutions of
differential equations}

Using different approaches for nonlinear differential equations, the
authors obtain different forms of the solutions and sometimes they
find the solutions in cumbersome forms. Interpretation and
understanding of such solutions are difficult and we believe, that
these solutions can be simplified. The consequence of the cumbersome
expressions is a risk of misprints. The solutions in cumbersome form
give an illusion that the authors indeed find new solutions. We meet
many such solutions of nonlinear differential equations especially
found by means of the Exp --- function method. Another consequence
is the fact, that the authors obtain the same solutions expressed in
different forms, because it is well known, that trigonometric and
hyperbolic expressions admit different representation.

The fifth error can be formulated as follows.

\emph{\textbf{Fifth error}. Expressions for solutions of nonlinear
differential equations are often not simplified.}

\emph{\textbf{Example 5a. }} \emph{Solutions of the Jimbo - Miva
equation \eqref{JM1} by \"{O}zi\c{s} and Aslan} \cite{Ozis02}.

Using the Exp - function method \"{O}zi\c{s} and Aslan \cite{Ozis02}
found some solitary wave solutions of Eq. \eqref{JM1}, but a number
of these solutions can be simplified. For example solution (30) in
\cite{Ozis02} is transformed by the following set of equalities
\begin{equation}\begin{gathered}
\label{JM2} u(x,y,z,t)=\frac{a_2\,e^{2\,\xi}+a_2\,b_1\,e^{\xi}+
b_0\,(a_2-4\,k)+b_0\,b_1\,(a_2-4\,k)\,e^{-\xi}}{e^{2\,\xi}+
b_1\,e^{\xi}+b_0+b_0\,b_1\,e^{-\xi}}=\\
\\
=\frac{\left(a_2\,e^{\xi}+b_0\,a_2\,e^{-\xi}-4\,b_0\,k\,e^{-\xi}\right)
\left(e^{\xi}+b_1\right)}{\left(e^{\xi}+b_0\,e^{-\xi}\right)\,\left(e^{\xi}+
b_1\right)}=a_2-\frac{4\,b_0\,k}{b_0+e^{2\,\xi}}.
\end{gathered}\end{equation}

It is seen, that the last expression is better for understanding,
then the solution by \"{O}zi\c{s} and Aslan.

\emph{\textbf{Example 5b. }} \emph{Solutions of the Konopelchenko -
Dubrovsky equation \eqref{KD} by Abdou} \cite{Abdou02}.

By means of the Exp - function method Abdou \cite{Abdou02} obtained
some solitary wave solutions of Eq. \eqref{KD}, but some solutions
can be simplified. For example solution (42) in \cite{Abdou02}
reduces to
\begin{equation}\begin{gathered}
\label{KD7}
u(x,t)=\frac{a_2\,e^{2\,\eta}+a_1\,e^{\eta}-\frac{a_{1}^2\,(a_2+2\,k^2)}
{2\,k^4}+\frac{a_{1}^3}{4\,k^4}\,e^{-\eta}+\frac{a_2\,a_1^4}{16\,k^8}\,
e^{-2\,\eta}}{\frac{a_{1}^4}{16\,k^8}\,e^{-2\,\eta}+
\,e^{2\,\eta}-\frac{a_{1}^2}{2\,k^4}}=\\
\\
=a_2+\frac{a_1\,\left(e^{\frac{\eta}{2}}-\frac{a_{1}}{2\,k^2}\,e^{-\frac{-\eta}{2}}\right)^2
}{\left(e^{{\eta}}-\frac{a_{1}^2}{4\,k^4}\,e^{{-\eta}}\right)^2}=a_2+
\frac{a_1}{\left(e^{\frac{\eta}{2}}+\frac{a_{1}}{2\,k}\,e^{-\frac{-\eta}{2}}\right)^2}.
\end{gathered}\end{equation}

\emph{\textbf{Example 5c}}. \emph{Solution of the modified Benjamin
- Bona - Mahony equation by Yusufo\v{g}lu}\cite{Yusufoglu01}
\begin{equation}
\label{BBM} u_t+u_x+\alpha\,u^2\,u_x+u_{xxx}=0.
\end{equation}

Using the travelling wave the author \cite{Yusufoglu01} obtained the
equation
\begin{equation}
\label{BBM1}
\beta^2\,U_{\xi\xi}+\frac13\,\alpha\,U^3+(1-\gamma)\,U=0.
\end{equation}
Yusufo\v{g}lu \cite{Yusufoglu01} found the solution of
Eq.\eqref{BBM1} in the form
\begin{equation}\begin{gathered}
u{(\xi)}=\frac{\frac{24b_0b_1(c-1)}{a_0a}e^{\xi}+a_0}
{\frac{24b_0b_1^2(c-1)}{a_0^2a}e^{2\xi}+b_1e^{\xi}+b_0+\frac{a_0^2a}{24b_1(c-1)}e^{-\xi}}
\label{BBM1a}.
\end{gathered}\end{equation}
Solution \eqref{BBM1a} has three arbitrary constants $a_0$, $b_0$
and $b_1$, but in fact this solution can be simplified using the set
of equalities
\begin{equation}\begin{gathered}
u{(\xi)}=\frac{\frac{24b_0b_1(c-1)}{a_0a}e^{\xi}+a_0}
{\frac{24b_0b_1^2(c-1)}{a_0^2a}e^{2\xi}+b_1e^{\xi}+b_0+\frac{a_0^2a}{24b_1(c-1)}e^{-\xi}}=\\
\\
=\frac{1}{\frac{a\,a_0}{24\,b_1(c-1)}e^{-\xi}+\frac{b_1}{a_0}\,e^{\xi}}
\label{BBM1b}=\frac{24\,b\,(c-1)}{a\,e^{-\xi}+24\,{b}^2\,(c-1)\,e^{\xi}}.
\end{gathered}\end{equation}
We can see, that solution \eqref{BBM1b} contains the only arbitrary
constant $b=\frac{b_1}{a_0}$.

Note that Eq.\eqref{BBM} can be easy transformed to the modified KdV
equation and this equation is not the BBM equation.

\emph{\textbf{Example 5d. }} \emph{Solutions of the fifth - order
KdV equation by Chun} \cite{Chun01}
\begin{equation}
\label{KdV5}
u_t+30\,u^2\,u_x+20\,u_x\,u_{xx}+10\,u\,u_{xxx}+u_{xxxxx}=0.
\end{equation}

Using the Exp - function method, Chun obtained seventeen exact
solutions of Eq. \eqref{KdV5} \cite{Chun01}. His first solution
(formula (25) in \cite{Chun01}) can be simplified to the trivial
solution, because
\begin{equation}\begin{gathered}
\label{KdV5a}
u_1=\frac{-k^4\,e^{\eta}+2\,k^2\,a_0-k^4\,b_{-1}\,e^{-\eta}}{2\,k^2\,
e^{\eta}-4\,a_0+2\,k^2\,b_{-1}\,e^{-\eta}}\equiv-\frac{k^2}{2},
\qquad \eta=k\,\left(x-\frac{7\,k^4}{4}\,t\right).
\end{gathered}\end{equation}

The sixth solution by Chun (formula (30) \cite{Chun01}) can be
simplified to the trivial solution as well
\begin{equation}\begin{gathered}
\label{KdVb} u_6=\frac{a_1\,e^{\eta}+\,a_1\,b_0+a_1
\,b_{-1}\,e^{-\eta}}{ e^{\eta}+b_0+\,b_{-1}\,e^{-\eta}}\equiv a_1,
\qquad \eta=k\,\left(x+\omega\,t \right).
\end{gathered}\end{equation}

The thirteenth (as well as the fourteenth and the fifteenth)
solution (formula (69) in paper \cite{Chun01}) is also constant as
we can see from the equality
\begin{equation}\begin{gathered}
\label{KdVc} u_{13}=\frac{a_1\,e^{2\,\eta}+a_1\,b_1\,e^{\eta}}{
b_1\,e^{2\,\eta}+b_1^{2}\,e^{-\,\eta}}\equiv\frac{a_1}{b_1}, \qquad
\eta=\left(k\,x+\omega\,t \right).
\end{gathered}\end{equation}

We recognized, that among seventeen solutions presented by Chun in
his paper twelve solutions (25), (27), (28) (30), (49), (51) (53),
(55), (57), (69), (70) and (71) satisfy the fifth-order KdV
equation, but solutions (25), (30), (69), (70) and (71) are trivial
ones (constants) and (51) is not the solitary wave solution. Six
solutions (25), (27), (49), (53), (55) and (57) are solitary wave
solutions, but these ones are known and can be found by means of
other methods.

\emph{\textbf{Example 5e. }} \emph{Solutions of the improved
Boussinesq equation by Abdou and co -authors} \cite{Abdou07b}.

Using the Exp - function method Abdou and his co - authors looked
for the solitary wave solutions of the improved Bouusinesq equation
\begin{equation}\begin{gathered}
\label{ASE1} u_{tt}-u_{xx}-u\,u_{xx}-(u_{x})^2-u_{xxtt}=0.
\end{gathered}\end{equation}
In the travelling wave Eq.\eqref{ASE2} takes the form
\begin{equation}\begin{gathered}
\label{ASE1a}(w^2-k^2-k^2\,u)\,u_{\eta\eta}-k^2\,u_{\eta}^2-k^2\,w^2\,u_{\eta\eta\eta\eta}=0.
\end{gathered}\end{equation}
The authors found several solutions of Eq.\eqref{ASE1a}. One of them
takes the form
\begin{equation}\begin{gathered}
\label{ASE2} u(\eta)=\frac{\left( {a_1}\,{e^{\eta}}-{\frac {{b_0}\,
\left( 5\,{a_1}\,{k}^{ 2}+{a_1}+6\,{k}^{2} \right)
}{{k}^{2}-1}}+\frac{a_1\,b_0^2}{4}{e^{-\eta}} \right)} {\left(
{e^{\eta}}+{b_0}+\frac{b_0^2}{4}{e^{- \eta}} \right)}, \quad
\eta=k\,x+t\,\sqrt{-\frac{a_1+1}{k^2-1}}.
\end{gathered}\end{equation}
However this solution can be presented as
\begin{equation}\begin{gathered}
\label{ASE3}u(\eta)={a_1}-{\frac {6\,{b_0}\,{k}^{2} \left( {a_1}+1
\right) }{
 \left( {k}^{2}-1 \right)  \left( {e^{\frac{\eta}{2}}}+\frac{b_0}{2}
 \,{e^{-\frac{\eta}{2}}} \right) ^{2}}}.
\end{gathered}\end{equation}
Solution \eqref{ASE3} can be obtained by many methods.

We have often observed the fifth error studying the application of
the Exp - function method.

Other examples of the fifth error are given in our recent papers
\cite{Kudryashov09a, Kudryashov09b}.

\section{Sixth error: some authors do not check solutions of
differential equations}

Using the computer methods to look for the exact solutions of
nonlinear differential equation we cannot remove our knowledge, our
understanding and our attention to the theory of differential
equations. Neglecting to do this we can obtain different mistakes in
finding the exact solutions of nonlinear differential equations.

These mistakes lead to numerous misprints in the final expressions
of the solutions and sometimes to the fatal errors. A need of
verifying the obtained solutions is obvious and it does not take
much time with the help of the modern computer algebra systems. We
have to substitute the found solution  into the equation and to
check that the solution satisfies it.

The sixth error is formulated as follows.

\emph{\textbf{Sixth error}. Some authors do not check the obtained
solutions of nonlinear differential equations.}

\emph{\textbf{Example 6a}}.  \emph{"Solutions" of the Burgers
equation by Soliman} \cite{Soliman01}.

Soliman \cite{Soliman01} found three "solutions" of the Burgers
equation \eqref{B}
\begin{equation}
\label{B2}
u_1(\xi)=\frac{c}{\alpha}+\frac{2\,\nu}{\alpha}\,\tanh{\{x-c\,t\}},
\end{equation}
\begin{equation}
\label{B2a}
u_2(\xi)=\frac{c}{\alpha}+\frac{c^2}{2\,\alpha\,\nu}\,\coth{\{x-c\,t\}},
\end{equation}
\begin{equation}
\label{B2b}
u_3(\xi)=\frac{c}{\alpha}+\frac{2\,\nu}{\alpha}\,\tanh{\{x-c\,t\}}+
\frac{c^2}{8\,\alpha\,\nu}\,\coth{\{x-c\,t\}}.
\end{equation}
However all these solutions are incorrect and do not satisfy
Eq.\eqref{B}. The solution of Eq.\eqref{B} can be found by many
methods and can be presented in the form
\begin{equation}
\label{B3}
u_4(\xi)=\frac{c}{\alpha}-\frac{2\,\nu}{\alpha}\,\tanh{\{x-c\,t-x_0\}}.
\end{equation}
where $x_0$ is arbitrary constant.

\emph{\textbf{Example 6b. }} \emph{Solution of the foam drainage
equation by Bekir and Cevikel} \cite{Bekir}
\begin{equation}
\label{FD} u_t+\frac{\partial}{\partial
x}\left(u^2-\frac{\sqrt{u}}{2}\,\frac{\partial u}{\partial
x}\right)=0.
\end{equation}

Eq.\eqref{FD} was studied in \cite{Bekir} by means of the tanh -
coth method by Bekir and Cevikel \cite{Bekir}. The authors proposed
the method to obtain "new travelling wave solutions".

Using the travelling wave
\begin{equation}
\label{FD1} u(x,t)=u(\xi), \qquad \xi=k\,(x+c\,t),
\end{equation}
they obtained the first - order equation in the form
\begin{equation}
\label{FD2}
c\,k\,u+k\,\left(u^2-\frac{k}{2}\,\sqrt{u}\,u_{\xi}\right)=0,
\end{equation}
and obtained three solutions
\begin{equation}\begin{gathered}
\label{FD3} u_1=k^2\,\tanh^2{\left(k(x-k^2\,t)\right)},\\
u_2=k^2\,\coth^2{\left(k(x-k^2\,t)\right)},\\
u_3=k^2\left(\,\tanh{\left(k(x-4\,k^2\,t)\right)}+
\,\coth{\left(k(x-4\,k^2\,t)\right)}\right)^2.
\end{gathered}\end{equation}

The solution of Eq.\eqref{FD2} can be found using the transformation
$u=v^2$. For function $v(\xi)$ we obtain the Riccati equation in the
form
\begin{equation}
\label{FD4} c+v^2=k\,v_{\xi}.
\end{equation}
Solution of Eq.\eqref{FD4} can be written as
\begin{equation}
\label{FD5} v(\xi)=-k\,\tanh(\xi-\xi_0), \qquad \xi=k\,x+k^3\,t.
\end{equation}
We can see there are the misprints in solutions of Eq.\eqref{FD2}

\emph{\textbf{Example 6c.}}  \emph{"Solutions" of the Fisher
equation by \"{O}zi\c{s} and K\"{o}ro\v{g}lu} \cite{Ozis01}
\begin{equation}
\label{F} u_{t}-u_{xx}-u\,(1-u)=0.
\end{equation}

Using the Exp-function method \"{O}zi\c{s} and K\"{o}ro\v{g}lu
\cite{Ozis01} found four "solutions" of Eq. \eqref{F}. These
"solutions" were given in the form
\begin{equation}\begin{gathered}
\label{F1}
u_{(1)}=\frac{(1-2\,k^2)\,b_{-1}}{{b_0}\,\exp{(\eta)}+b_{-1}},
\qquad \eta=k\,x+w\,t,
\end{gathered}\end{equation}

\begin{equation}\begin{gathered}
\label{F2}
u_{(2)}=\frac{(1-8\,k^2)\,b_{-1}}{{b_1}\,\exp{\left(2\,\eta\right)}+b_{-1}},\qquad
\eta=k\,x+w\,t,
\end{gathered}\end{equation}

\begin{equation}\begin{gathered}
\label{F3}
u_{(3)}=\frac{b_0-2\,k^2\,b_{-1}\,\exp{\left(-\eta\right)}}{{b_0}+{b_{-1}}
\exp{\left(-\eta\right)}}, \qquad \eta=k\,x+w\,t.
\end{gathered}\end{equation}

\begin{equation}\begin{gathered}
\label{F3a} u_{(4)}={\frac {{\it b_{1}}\,{{\rm
e}^{\eta}}-8\,{k}^{2}{\it b_{-1}}\,{{\rm e}^{-\eta}}}{ {\it
b_{1}}\,{{\rm e}^{\eta}}+{ b_{-1}}\,{{\rm e}^{-\eta}}}}, \qquad
\eta=k\,x+w\,t.
\end{gathered}\end{equation}

However all these expressions do not satisfy equation \eqref{F}. We
can note this fact without substituting solutions \eqref{F1} -
\eqref{F3a} into Eq. \eqref{F}, because a true solution of Eq.
\eqref{F} has the pole of the second order, but all functions
\eqref{F1} - \eqref{F3a} are the first order poles and certainly by
substituting expressions \eqref{F1} - \eqref{F3a} into equation
\eqref{F} we do not obtain zero.

\emph{\textbf{Example 6d.}}  \emph{"Solutions" of the modified
Benjamin - Bona - Mahony equation \eqref{BBM} found by Yusufoglu}
\cite{Yusufoglu01}.

The general solution of Eq.\eqref{BBM1} can be found via the Jacobi
elliptic function, but the author tried to obtain some solitary
solutions by means of the Exp - function method using the travelling
wave $U(\xi)$,    $\xi=\beta\,(x-\gamma\,t)$. Some of his solutions
are incorrect. For example the "solution" (formula
(3.12)\cite{Yusufoglu01})
\begin{equation}
\label{BBM2}
U(\xi)=\sqrt{\frac{3(c-1)}{\alpha}}\,\frac{\exp{\{\frac{\beta(x-\gamma\,t)}{2}\}}-
\frac{b_0}{2\,b_1}\,\exp{\{-\frac{\beta(x-\gamma\,t)}{2}}\}}
{\exp{\{\frac{\beta(x-\gamma\,t)}{2}\}}+\frac{b_0}{2\,b_1}\,
\exp{\{-\frac{\beta(x-\gamma\,t)}{2}}\}}
\end{equation}
do not satisfy Eq.\eqref{BBM1}.

The solution of Eq.\eqref{BBM1} takes the form
\begin{equation}
\label{BBM3} U(\xi)=\pm\,{\frac {6\,\beta\, }{\sqrt {-6\,\alpha
}}}\,\tanh \left( \xi-{\xi_0} \right) ,\qquad
\xi=\beta\,(x\,+\,\beta^2\,t - t),\quad \gamma=1-\beta^2
\end{equation}
This solution can be obtained by using different methods.

\emph{\textbf{Example 6e.}}  \emph{"Solutions" of the Burgers -
Huxley equation by Chun} \cite{Chun08a}.

Chun \cite{Chun08a} applied the Exp - function method to obtain the
generalized solitary wave solutions of the Burgers  --- Huxley
equation
\begin{equation}
  u_t+\alpha\,u\,u_x-u_{xx}=\beta\,u(1-u)(u-\gamma).\label{BHE1}
\end{equation}
In the travelling wave the author looked for the solution of
equation
\begin{equation}
  \omega\,u_{\eta}+\alpha\,u\,u_{\eta}-u_{\eta\eta}=\beta\,u(1-u)(u-\gamma),\label{BHE2}
\end{equation}
where $\eta=k\,x+\omega\,t$. Some of the solutions by Chun are
incorrect. In particular, "solution" (25) in \cite{Chun08a}
\begin{equation}
u(eta)=\frac{\gamma\,b_{-2}\,e^{-2\,\eta}}{e^{2\,\eta}+b_{-2}\,e^{-2\,\eta}},
\qquad
\eta=\frac{\gamma}{4}\,x+\frac{\gamma(1-\gamma)}{2}\,t\label{BHE3}
\end{equation}
do not satisfy Eq.\eqref{BHE2}. The solutions of Eq.\eqref{BHE1}
were found in \cite{Efimova04}.

\emph{\textbf{Example 6f.}}  \emph{"Solutions" of the Benjamin -
Bona - Mahony - Burgers equation by El-Wakil, Abdou and Hendi}
\cite{Elwakil08a}.

El-Wakil and co-authors \cite{Elwakil08a} using the Exp - function
method looked for the solitary wave solutions of the BBMB equation
\begin{equation}
 u_t-u_{xxt}-\alpha\,u_{xx}+u\,u_x+\beta\,u_{x}=0.\label{BBMB1}
\end{equation}
Taking the travelling wave $\eta=k\,x+c\,t$ the authors searched for
the solution of the equation
\begin{equation}
 c\,u_{\eta}-c\,k^2\,u_{\eta\eta\eta}-\alpha\,k^2\,u_{\eta\eta}
 +\beta\,k\,u_{\eta}+k\,u\,u_{\eta}=0.\label{BBMB2}
\end{equation}
Some of solutions by El-Wakil and co-authors are incorrect. In
particular, expression (18) in \cite{Elwakil08a}
\begin{equation}\begin{gathered}
u(\eta)=\frac{\left( -{\frac { \left( 6\,{b_0}\,{k}^{2}+{k}^{2}{
a_0}-{a_0} \right)}{{ b_0} \left( 1+5\,{k}^{2} \right) }}
{e^{\eta}}+{a_0}-\frac14\,{\frac {{b_0}\, \left( 6\,{
b_0}\,{k}^{2}+{k}^{2}{a_0}-{a_0} \right) }{1+5\,{k}^{2}}}
{e^{-\eta}}\right)} { \left( {e^{\eta}}+{b_0}+
\frac{{{b_0}}^{2}}{4}\,{e^{-\eta}} \right)} , \\
\\
\eta=k\,x-{\frac {k \left( {b_0}+{a_0} \right) }{{b_0}\,\left(
1+5\,{k}^{2} \right)}}\,t \label{BBMB3}
\end{gathered}\end{equation}
do not satisfy Eq.\eqref{BBMB2}.

The solutions of Eq.\eqref{BBMB2} can be found taking the different
methods into account. Note that integrating Eq.\eqref{BBMB2} with
respect to $\xi$ we have
\begin{equation}
u_{\eta\eta}+\frac{\alpha}{c}\,u_{\eta}-\frac{1}{2\,c\,k}\,u^2-\frac{c+\beta\,k}{c\,k^2}\,u
-C_2=0,\label{BBMB4}
\end{equation}
where $C_2$ is an arbitrary constant. Eq.\eqref{BBMB4} coincides
with the KdV - Burgers equation in the travelling wave. The solitary
wave solutions of this equation are given above in section 1. The
general solution of Eq.\eqref{BBMB4} can be found by analogue with
the general solution of Eq.\eqref{SMM4a} \cite{Kudryashov09b}.

\emph{\textbf{Example 6g.}}  \emph{"Solutions" of the Klein ---
Gordon equation with quadratic nonlinearity by Zhang}
\cite{Zhang08a}.

The Exp - function method was used by Zhang to obtain the
generalized solitonary solutions of the Klein - Gordon equation with
the quadratic nonlinearity
\begin{equation}
 u_{tt}-u_{xx}+\beta\,u-\gamma\,u^2=0.\label{KG1}
\end{equation}
The author \cite{Zhang08a} applied the travelling wave $u=U(\eta)$
$\eta=k\,x+w\,t$ and searched for the solution of the equation
\begin{equation}
(w^2-k^2\,\alpha^2)\,U_{\eta\eta}+\beta\,U-\gamma\,U^2=0.\label{KG2}
\end{equation}
At least two solutions of Eq.\eqref{KG2} by Zhang \cite{Zhang08a}
are incorrect and do not satisfy Eq.\eqref{KG2}
\begin{equation}\begin{gathered}
U_1={\frac {\beta}{\gamma}}-\frac{3\,{b_0}\,\beta}{{\gamma}{\left(
{b_1}\,{e^{k\,x+ w\,t}}+{b_0}+\,{\frac
{{{b_0}}^{2}}{4{b_1}}{e^{-k\,x-w\,t}}} \right)}},\quad
w=k\,\sqrt{\alpha^2+\frac{\beta}{k^2}}, \label{KG3}
\end{gathered}\end{equation}
\begin{equation}\begin{gathered}
U_2={\frac {\beta}{\gamma}}+\frac{{a_1}}{ \left(
{e^{kx+wt}}-\,{\frac {{a_1}\, {\gamma}^{2}}{3\,\beta}}+\,{\frac
{{{a_1}}^{2}{\gamma}^{2}}{{36\, \beta}^{2}}}\, {e^{-kx-wt}}\right)},
\quad w=k\,\sqrt{\alpha^2+\frac{\beta}{k^2}}. \label{KG4}
\end{gathered}\end{equation}
The solitary wave solutions of Eq.\eqref{KG2} was obtained in many
papers, because this equation coincides with KdV equation in the
travelling wave \cite{Kudryashov09b}. These solutions can be written
as the following
\begin{equation}\begin{gathered}
U^{(1)}=\frac{3\,\beta}{2\,\gamma}\,\left(1-\,\tanh^2(k\,\eta-k\,\eta_0)\right),
\quad \eta=k\,x\pm
\frac12\sqrt{\frac{4\,\alpha^2\,k^4-\beta}{k^2}}\,t, \label{KG5}
\end{gathered}\end{equation}
\begin{equation}\begin{gathered}
U^{(2)}=-\frac{\beta}{2\,\gamma}\,\left(1-3\,\tanh^2(k\,\eta-k\,\eta_0)\right),
\quad \eta=k\,x\pm
\frac12\sqrt{\frac{4\,\alpha^2\,k^4+\beta}{k^2}}\,t, \label{KG5}
\end{gathered}\end{equation}
where $\eta_0$ is an arbitrary constant.

\emph{\textbf{Example 6h.}}  \emph{"Solutions" of the Kuramoto -
Sivashinsky equation by Noor, Mohyud - Din and Waheed} \cite{Noor}.

Using the Exp - function method Noor, Mohyud-Din and Waheed
\cite{Noor} looked for the solutions of the Kuramoto---Sivashinsky
equation
\begin{equation}
u_t+uu_x+u_{xx}+u_{xxxx}=0.\label{NoorKS}
\end{equation}
These authors presented two expressions as the solutions of the
Kuramoto - Sivashinsky equation. Their first "solution" is written
as
\begin{equation}
u_1=-\frac{\left(a_0+a_{-1}\exp\left(-kx+\frac{\left(-a_0^2+
2ka_{-1}\right)kt}{a_{-1}}\right)\right)\left(a_0^2-ka_{-1}+
a_{-1}k^3\right)}{a_{-1}^2\exp\left(-kx+\frac{\left(-a_0^2+
2ka_{-1}\right)kt}{a_{-1}}\right)},\label{Noor1}
\end{equation}
where $a_0$, $a_{-1}$, $k$ are arbitrary constants (formula (26) in
\cite{Noor}).

The second "solution" by the authors takes the form
\begin{equation}
u_2=\frac{\left(a_2e^{2\eta}+a_0+a_{-2}e^{-2\eta}\right)a_{-2}}{b_{2}a_{-2}e^{2\eta}+
b_{-2}a_0+b_{-2}e^{-2\eta}}, \quad
\eta=kx-\frac{k\left(a_{-2}+8k^3b_{-2}+2kb_{-2}\right)}{b_{-2}}t,\label{Noor2}
\end{equation}
where  $a_0$, $a_{-2}$, $a_2$, $b_0$, $b_2$, $b_{-2}$, $k$ are
arbitrary constants (formula (34) in \cite{Noor}).

Here we have the sixth common error in the fatal form. Substituting
cited "solutions" \eqref{Noor1} and \eqref{Noor2}  into
\eqref{NoorKS}, we obtain that these "solutions" do not satisfy the
Kuramoto - Sivashinsky equation. Moreover, we cannot obtain zero
with any nontrivial values of the parameters. We guess, that using
the Exp - function method the authors \cite{Noor} did not solve the
system of the algebraic equations for the parameters.

The exact solutions of Eq.\eqref{NoorKS} were first found in
\cite{Kuramoto01}. These two solutions take the form
\begin{equation}\begin{gathered}
u^{(1)}={C_0}-{\frac {45\sqrt {19}}{361}}\,\tan \left(\frac{\sqrt {19}}{38}
\left( x-{C_0}\,t-{x_0} \right)\right)-\\
\\
-{\frac {15\sqrt {19} }{361}}\,  \tan^3\left( \frac{\sqrt
{19}}{38}\left( x-{C_0}\,t-{x_0} \right)\right)
^{3},\label{firstKSsolution}
\end{gathered}\end{equation}
\begin{equation}\begin{gathered}
u^{(2)}={C_0}-{\frac {135\sqrt {209}}{361}}\,\tanh \left(
\frac{\sqrt {209}}{38}\left( x-{C_0}\,t-{x_0} \right)  \right) +\\
\\
+{\frac {165\sqrt {209}}{361}}\, \tanh^3 \left( \frac{\sqrt
{209}}{38} \left( x-{ C_0}\,t -{x_0} \right)
\right),\label{secondKSsolution}
\end{gathered}\end{equation}
where $C_0$ and $x_0$ are arbitrary constants.

Many authors  tried to find new exact solutions of the Kuramoto -
Sivashinsky equation \eqref{NoorKS}. Some of authors \cite{Wazzan09,
WazwazKS, ChenKS} believe that they found new solutions but it is
not this case. Nobody cannot find new exact solutions of
Eq.\eqref{NoorKS}.

\section{Seventh error: some authors include additional arbitrary constants into solutions}

In finding the exact solutions of nonlinear differential equation
our goal is to find the general solution. But it is not possible in
many cases. In this situation we can try to look for the exact
solutions with the larger amount of arbitrary constants. However we
need to remember that the general solution of the equation of the
$n$ - th order can have only $n$ arbitrary constants. Nevertheless
we meet the papers, in which the authors present the solution of the
Riccati equation with two or even three arbitrary constants.
Moreover these authors say about the advantage of their approach in
comparison with other methods taking into consideration the amount
of arbitrary constants. Unfortunately in these cases there are two
possible variants for this kind of solutions. The first variant is
the author have obtained the solution with extra arbitrary constants
and the amount of these constants can be decreased by means of the
transformations. In the second variant the large amount of arbitrary
constants in the solution points out that the author have found
wrong solution.

So, we can formulate the seventh error as follows.

\emph{\textbf{Seventh error}.  Some authors include additional
arbitrary constants into solutions of nonlinear ordinary
differential equations.}

\emph{\textbf{Example 7a. }} \emph{Solution of the Riccati equation}
\begin{equation}
\label{RR} Y_{\xi}=\beta(1+Y^2).
\end{equation}

Using
\begin{equation}
\label{T1} Y(\xi)=-\frac{\psi_{\xi}}{\beta\,\psi},
\end{equation}
we obtain the linear equation of the second order
\begin{equation}
\label{T2} \psi_{\xi\xi}+ \beta^2\,{\psi}=0.
\end{equation}
The solution of equation \eqref{RR} takes the form
\begin{equation}
\label{T3}
Y(\xi)=i\,\frac{C_3\,e^{-i\,\beta\,\xi}-C_4\,e^{i\,\beta\,\xi}}
{C_3\,e^{-i\,\beta\,\xi}+C_4\,e^{i\,\beta\,\xi}.}
\end{equation}
At first glance we obtain more general solution \eqref{T3} of the
Riccati equation then \eqref{S}, but in fact these solutions are the
same. We can see, that one of the constants can be removed by
dividing the nominator and the denominator in solution \eqref{T3} on
$C_3$ (or on $C_4$). Denoting $C_5=C_4/C_3$ (or $C_5=C_3/C_4$) we
obtain the solution with the only arbitrary constant.

\emph{\textbf{Example 7b}}. \emph{Solution of the Sharma - Tasso -
Olver equation}.

In example 2f we have obtained the solution of Eq.\eqref{STO1} in
the form
\begin{equation}\begin{gathered}
u{(\xi)}=\frac{\sqrt{c}}{\,\sqrt{\alpha}}\,\,\frac{C_2\,
\exp{\left\{\frac{\xi\,\sqrt{c}}{\mu\,\sqrt{\alpha}}\right\}}-
C_3\,\exp{\left\{-\frac{\xi\,\sqrt{c}}{\mu\,\sqrt{\alpha}}\right\}}}
{C_1+C_2\,\exp{\left\{\frac{\xi\,\sqrt{c}}{\mu\,\sqrt{\alpha}}\right\}}+
C_3\,\exp{\left\{-\frac{\xi\,\sqrt{c}}{\mu\,\sqrt{\alpha}}\right\}}}
\label{STO6}.
\end{gathered}\end{equation}

However Eq.\eqref{STO1} has the second order, but solution
\eqref{STO6} contains tree arbitrary constants. Sometimes it is
convenient to leave three constants in solution \eqref{STO6}, but we
have to remember that solution \eqref{STO6} is not the general
solution and one of these constants is extra. We can remove one of
the constants as in example 7a and can write the general solution of
Eq.\eqref{STO1} in the form
\begin{equation}\begin{gathered}
u{(\xi)}=\frac{\sqrt{c}}{\,\sqrt{\alpha}}\,\,\frac{C_4\,
\exp{\left\{\frac{\xi\,\sqrt{c}}{\mu\,\sqrt{\alpha}}\right\}}-
C_5\,\exp{\left\{-\frac{\xi\,\sqrt{c}}{\mu\,\sqrt{\alpha}}\right\}}}
{1+C_4\,\exp{\left\{\frac{\xi\,\sqrt{c}}{\mu\,\sqrt{\alpha}}\right\}}+
C_5\,\exp{\left\{-\frac{\xi\,\sqrt{c}}{\mu\,\sqrt{\alpha}}\right\}}}
\label{STO7}.
\end{gathered}\end{equation}
Solution \eqref{STO7} of Eq.\eqref{STO1} is not worse then solution
\eqref{STO6}, but this solution is the general solution by
definition and all other solution can be found from this one.

\emph{\textbf{Example 7c}}. \emph{Solution \eqref{BBM1a} of the
modified Benjamin - Bona - Mahony equation
\eqref{BBM}}\cite{Yusufoglu01}
\begin{equation}\begin{gathered}
u{(\xi)}=\frac{\frac{24b_0b_1(c-1)}{a_0a}e^{\xi}+a_0}
{\frac{24b_0b_1^2(c-1)}{a_0^2a}e^{2\xi}+b_1e^{\xi}+b_0+\frac{a_0^2a}{24b_1(c-1)}e^{-\xi}}
\label{BBM2a}.
\end{gathered}\end{equation}

Solution \eqref{BBM2a} has three arbitrary constants $a_0$, $b_0$
and $b_1$, but in fact this solution can be simplified to the form
with one arbitrary constant as it was demonstrated in example 5c.

\emph{\textbf{Example 7d}}. \emph{"New exact solution" of the
Riccati equation by Dai and Wang}\cite{Dai}
\begin{equation}\begin{gathered}
\varphi_{\xi}+l_0+\varphi^2=0 \label{Ric}.
\end{gathered}\end{equation}

Dai and Wang \cite{Dai} had been looking  for the "new exact
solutions" of the Riccati equation \eqref{Ric} and obtained five
solutions. One of them (solution (17)\cite{Dai}) takes the form
\begin{equation}\begin{gathered}
\varphi{(\xi)}=\frac{\sqrt{-l_0}\,b_1\exp{(\sqrt{-l_0}\xi+\xi_0)}-\sqrt{-l_0}
a_{-1}\exp{(-\sqrt{-l_0}\xi-\xi_0)}}{b_1\,\exp{(\sqrt{-l_0}\xi+\xi_0)}+
\,a_{-1}\,\exp{-(\sqrt{-l_0}\xi-\xi_0)}} \label{Ric1}.
\end{gathered}\end{equation}
We can see, that solution \eqref{Ric1} has three arbitrary
constants, but two of them certainly can be removed.

\emph{\textbf{Example 7e.}}  \emph{Solutions of the Burgers - Huxley
equation by Chun} \cite{Chun08a}.

Using the Exp - function method Chun \cite{Chun08a} obtained the
generalized solitary wave solutions of the Burgers  --- Huxley
equation (see, Eq.\eqref{BHE6} in example 6e). One of his solution
(solution (32) in \cite{Chun08a}) takes the form
\begin{equation}\begin{gathered}
\label{BHE6}u(\eta)=\frac{\gamma(\gamma-1)(\gamma\,e^{{2\eta}}+a_1\,
e^{{\eta}}+a_0+a_{-1}\,e^{-\eta})}{\gamma(\gamma-1)(e^{{2\eta}}+b_1
e^{{\eta}}+a_{-1}e^{{-\eta}})+\gamma(a_1b_1+a_0-b_1^2)-a_0-a_1^2+a_1b_1}\\
\\
\eta=\frac{\gamma-1}{2}\,x+\frac{\gamma^2-1}{4}\,t
\end{gathered}\end{equation}

The Burgers --- Huxley equation is the second order one but solution
\eqref{BHE6} has three arbitrary constants. We suggested that this
solution is incorrect. We checked this solution and have convinced
that this solution does not satisfy Eq.\eqref{BHE2}.

Especially many solutions of equations with superfluous constants
were obtained by means of the Exp - function method. Such examples
can be found in many papers.

\section{Conclusion}

Let us shortly formulate the results of this paper. First of all we
have tried to give some classification of the common errors and
mistakes which occur in finding the solitary wave solutions of
nonlinear differential equations. We have illustrated these errors
using the examples from the recent publications in which the authors
presented the exact solutions of nonlinear differential equations.

We have tried to classify the wide - spread errors in finding the
exact solutions of nonlinear differential equations. Unfortunately
we have no possibility to consider all errors and all problems of
computer methods in finding the exact solutions. However we can note
that the authors of many papers do not analyze the solutions of
nonlinear differential equations that they obtain. Some errors are
mentioned above due to this reason. It is important to discuss the
solutions of nonlinear differential equations. We need to study how
a solution depends on the parameters of nonlinear differential
equation. It is useful to study the plots for the solutions of
differential equation. We are interested in the stable exact
solutions of nonlinear differential equations. These solutions are
useful for the investigation of physical processes.

In our paper we have discussed the solitary wave solutions of
nonlinear differential equations, but certainly we can find the same
errors in looking for the other exact solutions.

We believe the results of this paper will be useful for young people
who are going to study the exact solutions of nonlinear differential
equations. We also hope that the material of this paper will be
interesting for some referees.

\section {Acknowledgements}

This work was supported by the International Science and Technology
Center under Project B 1213.

\end{document}